\documentclass[aps,prb,reprint,groupedaddress,amsmath,amssymb]{revtex4-2}

\usepackage{mathtools}
\usepackage{bm}
\usepackage{graphicx}
\usepackage{booktabs}
\usepackage{subcaption}
\usepackage{hyperref}

\usepackage{amsthm}
\newtheorem{theorem}{Theorem}
\newtheorem{lemma}{Lemma}
\newcounter{asmpRE}
\usepackage{cleveref}% keep last
\crefname{figure}{Fig.}{Figs.}
\Crefname{figure}{Figure}{Figures}
\crefname{subfigure}{Fig.}{Figs.}
\Crefname{subfigure}{Figure}{Figures}
\crefname{equation}{Eq.}{Eqs.}
\Crefname{equation}{Equation}{Equations}
\crefname{section}{Sec.}{Secs.}
\Crefname{section}{Section}{Sections}
\crefname{subsection}{Sec.}{Secs.}
\Crefname{subsection}{Section}{Sections}
\crefname{subsubsection}{Sec.}{Secs.}
\Crefname{subsubsection}{Section}{Sections}
\crefname{table}{Table}{Tables}
\Crefname{table}{Table}{Tables}
\crefname{appendix}{Appendix}{Appendixes}
\Crefname{appendix}{Appendix}{Appendixes}
\crefname{subappendix}{Sec.}{Secs.}
\Crefname{subappendix}{Section}{Sections}
\crefname{subsubappendix}{Sec.}{Secs.}
\Crefname{subsubappendix}{Section}{Sections}
\crefname{theorem}{Theorem}{Theorems}
\Crefname{theorem}{Theorem}{Theorems}
\crefname{lemma}{Lemma}{Lemmas}
\Crefname{lemma}{Lemma}{Lemmas}

\begin{document}

\title{Quantum Material Generation via Coherent Hamiltonian Search}

\author{Jiang Yuhan}
\email{yuhan.jiang@wisc.edu}
\affiliation{Department of Physics, University of Wisconsin--Madison, Madison, WI 53706, USA}
\author{Matthew Otten}
\email{mjotten@wisc.edu}
\affiliation{Department of Chemistry, University of Wisconsin--Madison, Madison, WI 53706, USA}
\affiliation{Department of Physics, University of Wisconsin--Madison, Madison, WI 53706, USA}
\date{\today}

\begin{abstract}
Discovering new materials with targeted properties is a central goal of materials science, yet the space of candidates is vast. Data-driven generative models learn the distribution of known materials, but they extrapolate poorly, and a generated structure is not known to be stable. A material's properties follow from its Hamiltonian, and a quantum computer can calculate them from it. We introduce the quantum material generative algorithm (QMage), a coherent search, guided by the Hamiltonian, over crystalline configurations of atomic composition, unit cell structure, and atomic positions, for one whose calculated properties meet the targets to within a given error tolerance. The computation is a single coherent quantum circuit in which each configuration's Hamiltonian is block-encoded, its target eigenstate isolated and properties evaluated, and the satisfying configurations amplified by fixed-point search and measured. One framework covers many targets, from formation energy and band gaps to elasticity and conductivity. We design the algorithm for the mature fault-tolerant era and estimate its cost in logical qubits and Toffoli depth from a classically routine benchmark to targets where established methods are unreliable. The approach generates candidate materials from first principles, rather than generalizing from those known, and could enable materials discovery when the target properties are difficult to predict classically.
\end{abstract}

\maketitle

% ===== Main text (reading order) =====

\section{Introduction}\label{sec:intro}

Advances in materials have marked successive technological eras, from the Bronze
Age to present-day silicon-based electronics, and new materials with targeted
properties are continually sought.
The space of possible compositions and structures is vast.  Only a small fraction
of conceivable materials has been
characterized~\cite{hortonAcceleratedDatadrivenMaterials2025}.  Material
discovery has accordingly moved from experimental trial and error to
computation: density functional theory with high-throughput screening, open
materials databases, and, more recently, machine learning now propose
candidate materials at
scale~\cite{chengArtificialIntelligencedrivenApproaches2026}.
A recent deep-learning search predicted about 2.2 million candidate
structures, of which 736 have been independently realized in concurrent
experiments~\cite{merchantScalingDeepLearning2023}.  This paradigm depends on existing
data, and on density functional theory as the approximate source of the property
values from which it learns~\cite{hortonAcceleratedDatadrivenMaterials2025}.

Generative models learn the distribution of known materials and produce new
candidates~\cite{debreuckGenerativeAICrystal2025,chengArtificialIntelligencedrivenApproaches2026}.
They operate not on the material itself but on a constructed representation of it---descriptors, learned embeddings, or graph and voxel encodings---and that representation can lose information.  Graphs, for instance, are not invertible and lose geometry.  The learned
distribution interpolates within known materials but extrapolates poorly beyond
the training data.  A
generated structure offers no guarantee of stability.  Stability is determined
only afterward by a downstream first-principles calculation, and a high validity
score can reflect memorization rather than
discovery.  Where the model succeeds, it
is no more accurate than its training
data~\cite{debreuckGenerativeAICrystal2025}.  This limitation is not particular
to materials.  It follows from data-driven learning and applies to any design task approached
in this way.  Quantum machine learning~\cite{biamonteQuantumMachineLearning2017,jiangVersatileVariationalQuantum2025,ottenQuantumMachineLearning2020} and quantum generative models~\cite{benedettiGenerativeModelingApproach2019}, when trained on classical materials data, likewise depend on the training data and its representation.  Methods that encode classical data as quantum-state amplitudes also incur a state-preparation cost, which for arbitrary state vectors can require a number of two-qubit gates exponential in the number of qubits.  Quantum generative models have so far struggled to generate valid
outputs~\cite{smaldoneQuantumMachineLearning2025}.  On near-term hardware they
are limited by noise and scale, may underperform well-tuned classical models
without error correction~\cite{wayoDFTHybridClassical2025}, and face
trainability problems such as barren plateaus~\cite{mccleanBarrenPlateausQuantum2018}.  Neither the classical nor
the near-term quantum approach removes the dependence on data.

A material's properties are fixed by its Hamiltonian, and a quantum computer can
calculate them.  Quantum algorithms have begun to take this route, computing
from the Hamiltonian to search compositions, to optimize a molecule's structure,
to find the most stable configuration of a small system, to process ensembles
of molecular geometries, or to compute the free-energy difference between two
molecules~\cite{barkoutsosQuantumAlgorithmAlchemical2020,kassalQuantumAlgorithmMolecular2009,kosugiExhaustiveSearchOptimal2023,nishiyaFirstquantizedAdiabaticTime2024,cortesEfficientBlockEncodings2026,huangFullqubitAlchemistQuantum2026}.
The algorithm presented here searches crystalline
materials for one whose calculated properties meet the targets to within
a specified error tolerance.  The candidate space is spanned by atomic
composition, unit cell structure, and atomic positions.  Electronic states and their response to
perturbations govern a broad range of material properties, so calculating these
from the Hamiltonian covers many design targets within one fixed framework.
Among these properties
are the classically intractable ones, which a quantum computer can compute and a
classical one cannot.  Because the computation represents the quantum system directly rather than learning from data, which is biased toward known materials, it may in principle reach configurations inaccessible to classical methods, and possibly materials with no precedent among those known.  We show that this search can be run as a single coherent
circuit, building on and extending the established methods of fault-tolerant quantum
computation.  It holds the whole candidate space in superposition, evaluates each
structure's properties from its Hamiltonian, and amplifies the structures that meet the
targets.

This design follows the intuition that a quantum system calls for a
quantum-native computation, one that remains unbroken, with no intermediate measurement collapsing the state and no intervening
classical computation.  The whole search is then a single quantum
process, with the candidates compared and the matches amplified within one
evolution.  Instead of fitting the algorithm to the constraints of the early fault-tolerant
machines now anticipated~\cite{katabarwaEarlyFaultTolerantQuantum2024}, where logical qubits are scarce and circuits are kept
shallow, we design it for the mature regime in which it belongs.  There, logical
qubits can be provisioned as a computation requires, a single deep and fully
coherent circuit is feasible, and the main constraint is running time, not
the number of available qubits.

\section{Algorithmic Framework}\label{sec:framework}

\subsection{The algorithm at a glance}\label{sec:glance}

\begin{figure*}[t]
\centering
\begin{subfigure}{\textwidth}
\centering
\includegraphics[width=\textwidth]{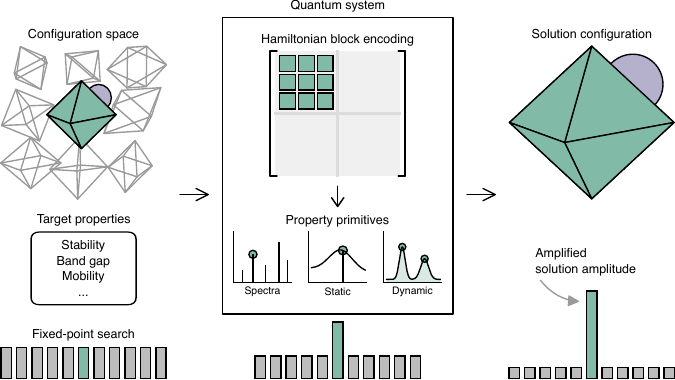}
\caption{}\label{fig:overview-a}
\end{subfigure}

\vspace{0.6em}
\begin{subfigure}{\textwidth}
\centering
\includegraphics[width=\textwidth]{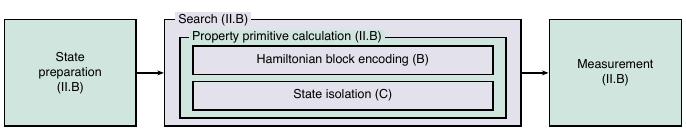}
\caption{}\label{fig:overview-b}
\end{subfigure}

\caption{Overview of QMage.
  (a)~The algorithm takes a space of configurations
  and a set of target properties (left) and returns a single configuration
  that meets the targets (right).  The quantum computer (center) block-encodes
  the Hamiltonian and evaluates the property primitives: spectral
  (eigenenergies), static (expectation values), and dynamical (response
  functions).  Fixed-point search amplifies the configurations that meet the
  targets.  The bar charts (bottom) show the configuration amplitudes growing
  from the uniform initial distribution to the amplified solution.
  (b)~Search contains the property primitive calculation
  (\cref{sec:definitions}), and that calculation contains the Hamiltonian
  block encoding (\cref{app:block-encoding}) and state isolation
  (\cref{app:isolation}).  The circuit begins with state preparation and ends
  with measurement.}%
\label{fig:overview}
\end{figure*}

Given a set of target material properties, the algorithm searches a space
of configurations and returns one that meets the targets, with a chosen success
probability (\cref{fig:overview}), when such configurations have support in
the prepared search space.  The computation is a single coherent quantum circuit
(\cref{fig:overview-b}), opened by a classical encoding that prepares the input
and closed by a classical decoding that reads the output.  Between those classical endpoints
the circuit passes through five stages: (i)~prepare a superposition over
material configurations~$x$; (ii)~block-encode each configuration's
Hamiltonian and isolate its target electronic eigenstate;
(iii)~evaluate the property primitives; (iv)~evaluate the flag function and amplify
the satisfying configurations by fixed-point search; (v)~measure, collapsing
onto a single configuration, then decode.  The configuration register is threaded through every stage. One and the same circuit acts on every configuration across the superposition.

We decompose material properties of interest into linear combinations of three classes of mathematical primitives: spectral (eigenenergies), static (expectation values), and dynamical (response functions), corresponding to the zeroth-, first-, and higher-order response of the target eigenstate.
The same framework can therefore evaluate different target properties expressed
in terms of these primitives.  Representative examples
include formation energy, phase stability, convex hull distance, and band
gaps (spectral); atomic forces, stress, magnetization, and charge density
(static); and phonons, elasticity, conductivity, and optical response
(dynamical)~\cite{hortonAcceleratedDatadrivenMaterials2025}.

Each stage uses a fixed unitary with the same circuit and iteration count
across the configuration superposition.  We choose the circuit size to
accommodate the most demanding configuration.  We evaluate properties
reversibly by computing the property values, using them to evaluate the flag
function for fixed-point search, and uncomputing the property registers while
preserving the configuration labels.

\subsection{Definitions and data flow}\label{sec:definitions}

\emph{Encoding and preparation.} Each configuration is a material configuration
$x = (\mathbf Z,\mathbf a,\mathbf R)$.  Here $\mathbf Z = \{Z_1,\dots,Z_N\}$
assigns one of $M$ allowed atomic species to each of $N$ sites (the atomic
composition), $\mathbf a = \{a,b,c,\alpha,\beta,\gamma\}$ specifies the unit
cell (lengths and angles), and
$\mathbf R = \{\mathbf R_1,\dots,\mathbf R_N\}$ lists the fractional atomic
coordinates. $\mathbf a$ and $\mathbf R$ are taken on discrete grids. All three registers are finite and bounded-precision, so the search space
is finite, of size $N_x = N_{\mathbf a}\,N_{\mathbf R}\,N_{\mathbf Z}$.  The
algorithm begins by preparing the initial superposition
\begin{equation}\label{eq:psi-init}
  \lvert\Psi_{\mathrm{init}}\rangle
  = \sum_x c_x\,\lvert x\rangle
    \Bigl(\sum_e c_e(x)\,\lvert e\rangle
      \otimes\lvert 0\dots 0\rangle_{\mathrm{anc}}\Bigr)^{\!\otimes N_P},
\end{equation}
where the material configuration state decomposes as
$\sum_x c_x\lvert x\rangle
 = \sum_{\mathbf Z,\mathbf a,\mathbf R} c_x\,
   \lvert\mathbf Z\rangle\lvert\mathbf a\rangle\lvert\mathbf R\rangle$.
Each configuration is paired with $N_P$ independent electronic states
(basis label~$e$, coefficients~$c_e$), one per property primitive, and their zero-initialized ancillas.  The coefficients
$c_x$ are either uniform ($1/\sqrt{N_x}$, an unbiased search) or set by prior
knowledge (a heuristic search).  Uniform preparation is elementary.  A
heuristic distribution uses standard state-preparation techniques~\cite{mottonenTransformationQuantumStates2005}.

\emph{Hamiltonian block encoding.} We define each configuration's Born--Oppenheimer electronic Hamiltonian in
\cref{app:hamiltonian} as a discrete, finite-cell operator,
$H(x) = T + U_{en} + V_{ee} + V_{nn}$
(kinetic, electron--nucleus, electron--electron~\cite{suFaultTolerantQuantumSimulations2021,babbushQuantumSimulationChemistry2019,nguyenQuantumComputingCorrosion2026}, and nucleus--nucleus
terms).  The electron--nucleus
interaction is given by the Hartwigsen--Goedecker--Hutter (HGH) separable
pseudopotential~\cite{hartwigsenRelativisticSeparableDualspace1998}, which leaves only the valence electrons and retains the usually omitted relativistic spin--orbit interaction.  Written
in a first-quantized dual basis~\cite{suFaultTolerantQuantumSimulations2021}, the Hamiltonian keeps the kinetic term diagonal
in momentum and the interactions in position.  All dependence on the cell
geometry is captured by the crystallographic metric tensor (or Gramian matrix)
$\mathbf G(\mathbf a)$, which maps fractional coordinates to physical energy.
This allows us to use the same block-encoding construction for all cells in
the search space.  A single electronic register, holding the largest number of active electrons, likewise serves every composition.  Following the linear-combination-of-unitaries (LCU)
construction~\cite{suFaultTolerantQuantumSimulations2021}, we build an exact
block encoding $\mathsf U_H(x)$ of the implemented operator $\widetilde H(x)$
and prove its validity in \cref{app:block-encoding}.

\emph{Property primitive calculation.} We then extend discrete adiabatic evolution~\cite{anLargeTimestepDiscretisation2025}
across the configuration superposition, isolating every configuration's target
eigenstate $\psi_n(x)$ with a single fixed unitary, block-diagonal in $x$, and
prove its convergence in \cref{app:isolation}, producing
$\lvert\Psi_{\mathrm{target}}\rangle
 = \sum_x c_x\lvert x\rangle\bigl(\lvert\psi_n(x)\rangle_e
   \otimes\lvert 0\rangle_{\mathrm{anc}}\bigr)^{\!\otimes N_P}$.  With the target eigenstate in hand, we express each property primitive in one of
three forms:
\begin{align}
  \text{spectral:}\quad
    & E_n = \langle\psi_n\lvert\hat H\rvert\psi_n\rangle,
    \label{eq:prim-spec}\\
  \text{static:}\quad
    & \langle\psi_n\lvert\hat V\rvert\psi_n\rangle,
    \quad \frac{\partial E_n}{\partial q}
    = \Bigl\langle\psi_n\Bigl\lvert
      \frac{\partial\hat H}{\partial q}
      \Bigr\rvert\psi_n\Bigr\rangle,
    \label{eq:prim-stat}\\
  \text{dynamical:}\quad
    & \langle\psi_n\lvert\hat V\,f(\hat H)\,\hat V
      \rvert\psi_n\rangle,
    \label{eq:prim-dyn}
\end{align}
where $\hat V$ is a generalized operator, $q$ is a generalized parameter (so that the static derivative gives forces and stresses by
the Hellmann--Feynman theorem~\cite{feynmanForcesMolecules1939}), and $f(\hat H)$ denotes the resolvent
$f(\hat H) = (\hat H - E_n - \textstyle\sum\omega - i\eta)^{-1}$, with
$\omega$ the driving frequencies and $\eta$ the broadening parameter.  The displayed dynamical form uses a single resolvent.  A response of order~$r$ is the $r$-th energy derivative $\partial^r E_n/\partial q^r$, which equals a static contribution plus a sum over the $r!$ permutations of $r$ nested resolvents.  Spectral, static, and dynamical primitives are then the zeroth, first, and higher orders of a single expansion.  Each primitive $P_j$, $j=1,\dots,N_P$, is computed by a unitary
$\mathsf U_{P_j}$ that writes its value into a dedicated register.  The spectral primitive reads
the target eigenenergy $E_n$ by qubitization and tapered quantum phase
estimation~\cite{lowHamiltonianSimulationQubitization2019,patelOptimalCoherentQuantum2026}.  It forms the qubitization walk of the block encoding and estimates $E_n$ into the energy register, the taper concentrating the value enough to remain coherent.  The spin--orbit terms are not self-inverse, but each is paired with its adjoint at the same coefficient in the linear combination, and that pairing is enough for a valid qubitization walk.  The static primitive estimates an expectation value by
coherent iterative amplitude estimation.  It block-encodes the operator and
shifts it so the expectation lies in $[0,1]$, then applies the non-destructive
projection onto the target eigenstate and the coherent iterative estimation
of~\cite{rallFasterCoherentQuantum2021}.  The dynamical primitive combines the two.  It uses the eigenenergy to shift the Hamiltonian and the same
amplitude estimation to read out, with quantum signal processing or a quantum singular value
transformation between the operators to realize the resolvent~\cite{lowOptimalHamiltonianSimulation2017,gilyenQuantumSingularValue2019}.  The result is summed over the permutation index.

\emph{Search and measurement.} The computed primitive values yield derived properties $A, B, C, \dots$, each compared to its target. The outcome is a bit, 1 if the constraint is met and 0 otherwise. We combine these bits by union~(OR), exclusion~(NOT), and intersection~(AND) into a single flag function $f(x)$.  We then amplify the flagged
(satisfying) configurations to a success probability of at least~$p$ by fixed-point
search~\cite{yoderFixedPointQuantumSearch2014}. The number of iterations is set from a lower bound on the solution density $\rho$ in the prepared state. In the uniform enumerated case, $\rho = N_{sol}/N_x$, with $N_{sol}$ the number of solution configurations.  For a heuristic search, $\rho$ is the sum of $\lvert c_x\rvert^2$ over the satisfying configurations.  Because each iteration computes and then uncomputes the
property evaluation, the search reshapes only the configuration amplitudes.
After $m$~iterations the pre-measurement state
$\lvert\Psi_{\mathrm{pre}}\rangle$ is $\lvert\Psi_{\mathrm{init}}\rangle$
with the configuration amplitudes amplified, $c_x \to c_x^{(m)}$, so that
$\sum_{x\,\in\,\mathrm{sol}}\lvert c_x^{(m)}\rvert^2 \ge p$.  Measurement
then collapses it to the final state
$\lvert\Psi_{\mathrm{final}}\rangle = \lvert x_{sol}\rangle\bigl(\sum_e
c_e(x_{sol})\lvert e\rangle\otimes\lvert 0\dots0\rangle_{\mathrm{anc}}
\bigr)^{\!\otimes N_P}$, a single solution configuration, with probability
at least~$p$.  The measured configuration is then decoded classically and handled by the standard materials procedures.  We
use the register discipline compute--use--uncompute.
Each primitive register and each search-body register is allocated, computed, used,
and uncomputed as a whole, so the configuration register is clean between
iterations and the $N_P$~primitives evaluate in parallel, each on its own
independently prepared electronic state.

We estimate the cost on three crystalline search tasks: stoichiometric lithium hydride, a classically routine benchmark; a solid-state tritium breeder, a classically hard target; and a three-dimensional quantum spin liquid, where established methods are unreliable.

\section{Resource Analysis}\label{sec:resources}

\begin{figure}[t]
\centering
\includegraphics[width=\columnwidth]{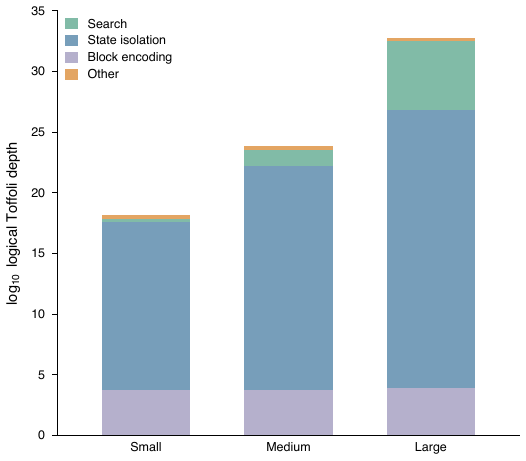}
\caption{Logical Toffoli-depth scaling.  Base-10 logarithm of the logical
  Toffoli depth at three illustrative problem scales of increasing size
  (Small, Medium, Large), at the inputs of \cref{tab:scaling-points}.
  The total depth factors into the number of fixed-point search iterations, state
  isolation, the block encoding, and the remaining minor factors, shown as stacked
  $\log_{10}$ contributions, so the bar height is the $\log_{10}$ of the total
  depth.  State isolation dominates at every scale, while the search factor
  grows with system size and exceeds the block encoding at the largest scale.}%
\label{fig:scaling}
\end{figure}

We quantify the circuit by two measures, its number of logical qubits and its logical Toffoli depth, first as asymptotic bounds in terms of a single error tolerance~$\epsilon$, then as explicit numerical estimates for the case studies.  All costs cover
the quantum computation only.  The number of qubits, additive over registers,
is dominated by the pairwise potential-energy arithmetic ancillas, of
order $N_P(N_e{+}N)^2\log(1/\epsilon)$, with $N_P$ the number of property
primitives, $N_e$ the largest number of active electrons, and $N$ the number
of atomic sites.  The depth is the
number of fixed-point-search iterations
$m=\mathcal O\!\bigl(\rho^{-1/2}\bigr)$, with $\rho$ the solution density, times a
search-iteration body comprising state isolation, the dominant property
primitive, and the primitive-to-property arithmetic.
Because all active primitives are evaluated within the same configuration
superposition, the search-iteration body takes the maximum over the active
primitive classes rather than their sum, so the depth does not grow with the
number of target properties.  The Toffoli depth is dominated by
the state-isolation term, the product of the number of search iterations~$\rho^{-1/2}$,
the isolation factor~$(\lambda/\Delta)^2\log^3(N_P\lambda/(\sqrt{\rho}(1-p)\Delta\epsilon))$, and the block-encoding
depth~(\cref{fig:scaling}), where $\lambda$ is the block-encoding normalization and $\Delta$ the spectral gap isolating the target eigenstate.  The normalization~$\lambda$ is
$\mathcal O(N_e/\epsilon^2)$, set by the kinetic-energy coefficients that
grow with the largest momentum the grid must resolve.
We give the full bounds in \cref{app:resource-estimation}.

Within our encoding and discrete adiabatic evolution, the dominant depth
factors are the normalization~$\lambda$ and the gap dependence of state
isolation.  Both are set by the methods used for these steps.
The normalization is set by the largest kinetic energy the grid must
resolve.  The first-quantized LCU block encoding is the lowest-cost
state of the art in the regimes of interest, since alternative encodings
lower one term but raise the total practical
cost~\cite{suFaultTolerantQuantumSimulations2021}.  The quadratic gap
dependence is what one fixed evolution serving every candidate requires,
since no candidate's gap is known in advance.  Prior knowledge of the
candidate configurations allows an evolution adapted to the gap, at the
optimal linear scaling in the inverse gap~\cite{rolandQuantumSearchLocal2002}.
Further improvements in the block encoding or state isolation could reduce
these contributions to the depth.

We start with the most basic target, the ground-state energy.  Given only the
Li:H stoichiometry, the search would return the rocksalt phase of lithium hydride,
the lightest ionic compound, at ${\sim}10^{2}$ logical qubits and
${\sim}10^{14}$ Toffoli depth.  Because classical
electronic-structure methods already determine such ground-state energies
accurately, this case is a benchmark.
The algorithm is intended for classically hard problems, where
established methods meet well-documented, case-specific limitations that
motivate an exact many-electron simulation.  The solid-state tritium breeder
(${\sim}10^5$~qubits, ${\sim}10^{19}$~depth) is a candidate lithium ceramic
to breed tritium fuel for fusion.  It targets correlated,
polaronic oxide-defect states that semilocal DFT delocalizes without a
Hubbard-$U$ or hybrid
correction~\cite{fallettaHubbardPolaronicDefect2022}.  The
three-dimensional topological quantum spin liquid
(${\sim}10^7$~qubits, ${\sim}10^{26}$~depth) is a magnet with no conventional
magnetic order even at zero temperature, a platform for topological quantum
computation.  It targets a frustrated quantum magnet in which quantum Monte
Carlo loses its sign-free property
once non-Kitaev interactions or fields are present, and
purification-based thermal tensor networks converge poorly near the ground
state~\cite{jahromiThermodynamicsThreedimensionalKitaev2021}.  We
detail the reductions and inputs for each case in \cref{app:case-study}.

\begin{figure}[!t]
\centering
\includegraphics[width=\columnwidth]{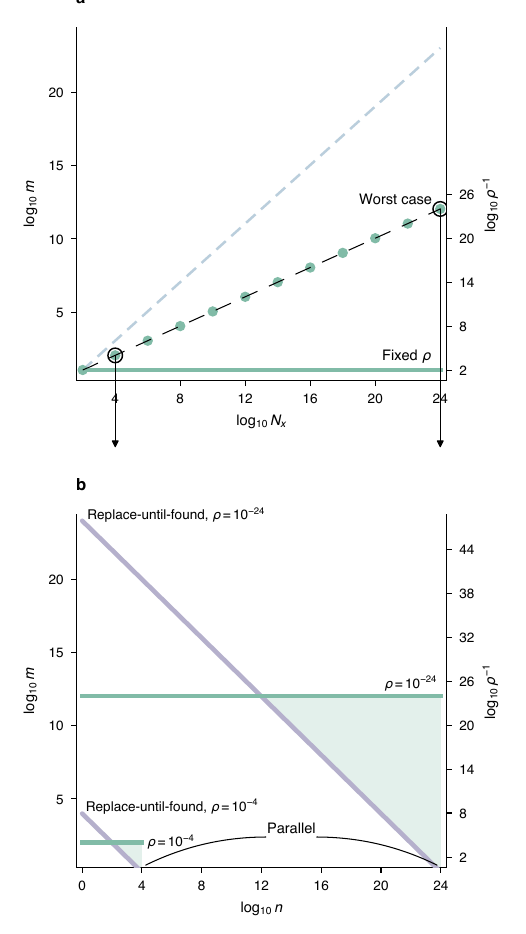}
\caption{Search-iteration scaling.  Success probability $p = 0.95$.
  (a)~The number of fixed-point search iterations~$m$ (left axis), with
  the corresponding inverse solution density~$\rho^{-1}$ (right axis),
  against the number of configurations~$N_x$ on base-10 logarithmic axes.
  At a fixed solution density, $m$ is flat in~$N_x$ (solid green line, here
  $\rho = 10^{-2}$).  For a single solution, where $\rho = 1/N_x$, $m$
  grows as $\sqrt{N_x}$ (filled green circles on a black dashed curve),
  sublinear in~$N_x$ against the linear reference (blue dashed line).
  Open circles mark the solution densities shown in panel~(b).
  (b)~$m$ and $\rho^{-1}$ against the number of devices~$n$ at
  $\rho = 10^{-4}$ and $\rho = 10^{-24}$.  Replace-until-found has depth
  $1/(n\rho)$ (purple).  Fixed-point search is independent of~$n$ (green).
  The shaded region is where the former's depth is less than the latter's,
  $n>\rho^{-1/2}$.  At $n=1/\rho$, replace-until-found reduces to the
  parallel case.}%
\label{fig:search-scaling}
\end{figure}

We consider the alternative of parallel fixed-input simulation,
one independent quantum simulation for each of the $N_x$ configurations, with the
satisfying ones selected by a final classical search over the
resulting values.  For a realistic search task this is not a viable
alternative.  $N_x$ grows exponentially with system size, so replicating the
simulation across all configurations demands hardware that cannot be built.
One can instead run $n$ devices, each evaluating one configuration at a time
and replacing it until a solution is found.  Relative to the fixed-point search,
this has sequential depth of order $1/(n\rho)$ rather than $\rho^{-1/2}$,
which is smaller when $n>\rho^{-1/2}$, and uses $n$ times as many qubits
(\cref{fig:search-scaling}b), and returns the property values of every
evaluated configuration, compared with one measured solution.
This is feasible only when $n$ is small and $\rho$ is large, where the search
depth is already modest.
Our algorithm instead carries out the search within a single circuit,
at the cost estimated above.  Its depth has no direct linear dependence
on~$N_x$, remaining flat at a fixed solution density~$\rho$ and bounded by
the worst-case single-solution scaling~$\sqrt{N_x}$
(\cref{fig:search-scaling}a), with $\rho$ discussed under \emph{Solution
density} in \cref{sec:re-case}.

\section{Discussion}\label{sec:discussion}

We write the algorithm for the mature fault-tolerant era, one in which error
suppression has become effectively scale-independent, so that logical qubits can
be provisioned as a computation requires and long coherent circuits remain
meaningful.  Moreover, physical qubits are inexpensive here.  Below the
fault-tolerance threshold, a modest increase in their number sharply lowers the
logical error rate~\cite{gottesmanIntroductionQuantumError2009}.
In this regime the largest computations are limited by their running
time rather than by the number of available qubits, and the circuit-size-versus-sample trade-offs of more constrained settings matter less~\cite{katabarwaEarlyFaultTolerantQuantum2024}.  The situation is
analogous, in broad outline, to mature classical computing, where the practical
concern is a computation's running time more than the number of available
components.  We therefore assume sufficient logical qubits for the ancilla
registers and sufficient error suppression to execute the full coherent search.

The principal limitation is the large Toffoli depth.  Improvements to the
search or to the underlying simulation methods could reduce this cost.
Our fixed-point search amplifies the
satisfying configurations within the superposition of all $N_x$ candidates, so
its cost depends on the solution density in that space.  The search requires
more iterations as the solution density decreases.  Its quadratic improvement
over unstructured classical search has optimal query
scaling~\cite{bennettStrengthsWeaknessesQuantum1997} and has also been used for
molecular structure optimization~\cite{kassalQuantumAlgorithmMolecular2009}.
A continuous-variable alternative would instead descend to
an optimum directly, without enumerating the space~\cite{catliEfficientQuantumOptimization2026}.  Its local cost is
set by the conditioning of the landscape and the dimension of the search, not by
$N_x$.  Such a search would be favorable where the landscape is smooth and solutions sparse; ours where the landscape is rugged or solutions are not rare.  Realizing the continuous search raises two foundational questions: how to re-express
the categorical atomic composition as a continuous, differentiable variable, and whether the search retains a
usable spectral gap on the resulting landscape.

Improvements to the underlying methods could also reduce the cost.  The
dominant depth factors depend on the block-encoding normalization and the
gap dependence of state isolation (\cref{sec:resources}).  Advances in such
methods have reduced the estimated costs of fault-tolerant applications by
orders of magnitude over the past
decade~\cite{babbushGrandChallengeQuantum2025}.

Within the same architecture, two natural constructions are not carried out here.
Our algorithm evaluates properties at zero temperature, isolating a
single pure target eigenstate.  Extending it to finite temperature would replace
that pure-state isolation step with preparation of the thermal Gibbs state at
the target temperature~\cite{poulinSamplingThermalQuantum2009}, which would
require further study for coherent superposition inputs.  We give the full
construction for the spectral primitive.  For the static and dynamical
primitives, we describe the methods and estimate their resources, while
leaving explicit circuit constructions for future work.

Finally, we build the framework for crystalline materials.  Non-crystalline materials
can be accommodated approximately within the first-quantized cell by vacuum
padding or spherical truncation~\cite{suFaultTolerantQuantumSimulations2021},
while a native formulation would use real-space grids with open boundary
conditions, a different architecture reserved for future work.  More broadly,
the architecture itself is not specific to materials.  A first-principles
Hamiltonian, property primitives, and a coherent configuration search transfer
in principle to other first-principles design problems.

We presented an algorithmic framework for searching crystalline configurations
by calculating their properties from the Hamiltonian within a single coherent
quantum circuit.  We estimated its resource requirements for cases ranging
from lithium hydride to targets where established methods are unreliable.
With sufficiently capable fault-tolerant quantum computers, this approach
could enable searches for materials whose properties are difficult to
calculate classically.

\begin{acknowledgments}
This work was supported by funding from the NSF QLCI for Hybrid Quantum
Architectures and Networks (NSF award 2016136).
\end{acknowledgments}

\section*{Author Contributions}
M.~O. and J.~Y. conceptualized the project. Both authors developed the
algorithm, performed the resource calculations, and wrote the paper.
The authors used AI tools to assist with manuscript drafting and editing,
literature searches, and review of mathematical derivations and resource
estimates. The authors reviewed and revised the resulting material and take
responsibility for the final manuscript.

% ===== Appendices =====
\appendix
\crefalias{section}{appendix}

\section{Hamiltonian}\label{app:hamiltonian}

We work with a Born--Oppenheimer electronic Hamiltonian defined for each material
configuration $x=(\mathbf Z,\mathbf a,\mathbf R)$ introduced in the main text.
The Hamiltonian is
\begin{equation}\label{eq:hamiltonian}
  H(x) = T(\mathbf a)
       + U_{en}(\mathbf R,\mathbf Z,\mathbf a)
       + V_{ee}(\mathbf a)
       + V_{nn}(\mathbf R,\mathbf Z,\mathbf a),
\end{equation}
with $T$ the kinetic energy, $U_{en}$ the electron--nucleus interaction,
$V_{ee}$ the electron--electron repulsion~\cite{babbushQuantumSimulationChemistry2019}, and $V_{nn}$ the nucleus--nucleus
repulsion.
The nuclei are fixed classical point charges for each
configuration.  Only the electrons are dynamical variables.

We express all geometric quantities through the crystallographic metric
tensor $\mathbf G(\mathbf a)$, the $3\times 3$ matrix of cell-vector inner products
whose diagonal components are $a^2,b^2,c^2$ and whose off-diagonal components are
$ab\cos\gamma$, $ac\cos\beta$, $bc\cos\alpha$.  Pairwise distances follow the
minimum-image convention. Given a fractional coordinate difference
$\Delta\mathbf r_{ij}$, the physical distance is
\begin{equation}\label{eq:min-image}
  d_{ij} = \min_{n\in\{-1,0,1\}^3}
    \sqrt{(\Delta\mathbf r_{ij}+n)^T\,\mathbf G\,
          (\Delta\mathbf r_{ij}+n)}\,;
\end{equation}
analogous expressions define $d_{iI}$ and $d_{IJ}$.  This finite neighborhood search is what the algorithm evaluates.  Component-wise modular wrapping is recovered only in the special case of an
orthogonal cell.  These conventions define a discrete,
finite-cell Hamiltonian whose individual terms we specify below.

\emph{Assumptions.}
(i)~Born--Oppenheimer: we treat nuclei as fixed classical point charges for
each configuration.
(ii)~A single $N_e$-particle first-quantized register serves the whole bounded
search, with $N_e$ the largest number of active electrons it must hold.  The configuration $x$ appears only through
the coefficients of $H$.  The register is therefore
configuration-independent.
For a configuration whose number of valence electrons is below $N_e$, we place the
$N_e - \sum_I Z_{\mathrm{ion},I}$ inactive padding electrons in distinct orbitals and make them non-interacting, removing their $T$, $U_{en}$, and
$V_{ee}$ contributions and satisfying Pauli exclusion without shifting $E_0$.  Charge neutrality holds because the active electrons balance
$\sum_I Z_{\mathrm{ion},I}$.
(iii)~The electron--electron and nucleus--nucleus interactions use
minimum-image distances with a maximum resolution limit (defined below).  The algorithm defines this discrete, finite-cell,
truncated-Coulomb Hamiltonian and block-encodes it up to a controlled
sampling error.  This Hamiltonian approximates full periodic (Ewald / lattice-sum)
electrostatics in the standard finite-cell sense set by the assumptions above.
Allowed cells are Niggli-reduced, hence Minkowski-reduced, and form a shortest-vector basis.  For a
Minkowski-reduced basis the minimum image lies within the
$n\in\{-1,0,1\}^3$ neighborhood~\cite{bernsteinInvertibleSevendimensionalDirichlet2023}.
Long-range tails and finite-size effects lie outside the finite-cell scope.
(iv)~The real-space grid has $N_{\mathrm{grid}}$ points per cell.  For a
nonorthogonal cell the grid spacing is set by the shortest grid lattice vector
(determined by the metric tensor), rather than by $[\det\mathbf G]^{1/2}$ alone.
(v)~The kinetic energy is non-relativistic ($\propto k^2$).  Relativistic
effects are captured by the pseudopotential.

\emph{Basis and regularization.}
We describe the electrons in a first-quantized dual basis.
The kinetic energy in a general cell is
\begin{equation}\label{eq:kinetic}
  T = 2\pi^2 \sum_{i=1}^{N_e}
      \mathbf k_i^{\,T}\,\mathbf G^{-1}(\mathbf a)\,\mathbf k_i\,,
\end{equation}
where the reciprocal metric tensor $\mathbf G^{-1}$ gives the kinetic energy
for a general (nonorthogonal)
cell (Eq.~(5) of~\cite{berryQuantumSimulationRealistic2024}).
We define the register on a fixed grid of fractional
coordinates.  It is therefore
configuration-independent.
All dependence on the cell (via $\mathbf G$), $\mathbf R$, and $\mathbf Z$
appears only through $H$'s coefficients.  A single register consequently
represents the electronic state for every configuration, the property that the
coherent search requires.  Plane waves are periodic by construction and match
crystalline boundary conditions.

We store the electronic state in the real-space (dual) basis so that
potentials, being functions of distance, are computed directly from coordinates
via $\mathbf G$.  The kinetic term, the only momentum-diagonal term, is
recovered by a quantum Fourier transform using $\mathbf G^{-1}$.  Each
Hamiltonian term is thereby evaluated in the representation where it is
diagonal.

Three regularizations remove Coulomb singularities.
We bound the electron--electron and nucleus--nucleus interactions:
\begin{align}
  V_{ee} &= \sum_{i<j}^{N_e}
    \min\!\bigl(\tfrac{1}{d_{ij}},\, V_{\mathrm{max}}\bigr),
    \label{eq:vee}\\
  V_{nn} &= \sum_{I<J}^{N}
    \min\!\bigl(\tfrac{Z_{\mathrm{ion},I} Z_{\mathrm{ion},J}}{d_{IJ}},\,
                 Z_{\mathrm{ion},I} Z_{\mathrm{ion},J}\, V_{\mathrm{max}}\bigr).
    \label{eq:vnn}
\end{align}
Here $V_{\mathrm{max}}$, the maximum resolution limit, is the inverse of the shortest
distinct grid-point separation, which is set by the shortest grid lattice
vector and thus determined by the metric.  For a cubic cell this reduces to
$V_{\mathrm{max}} \approx N_{\mathrm{grid}}^{1/3}\!/[\det\mathbf G]^{1/6}$.
The bound removes the $1/d$ divergence at zero separation ($d\to 0$),
assigning the closest-distinct-grid-point value.  It leaves every distinct-pair
value unchanged, and the deviation from the continuum kernel vanishes
as the grid is
refined.
The electron--nucleus singularity is removed by the non-singular $\mathrm{erf}$
local term of the pseudopotential, described next.

The dual-basis discretization converges as the grid is
refined.
We control the pseudopotential approximation and the Coulomb truncation
separately rather than through a single variational bound on the full $H$.

\emph{Pseudopotential.}
The electron--nucleus interaction $U_{en}=V_{\mathrm{loc}}+\hat V_{\mathrm{nl}}$
is given by the Hartwigsen--Goedecker--Hutter (HGH) dual-space Gaussian
pseudopotential~\cite{hartwigsenRelativisticSeparableDualspace1998}.
The local part is
\begin{equation}\label{eq:vloc}
\begin{split}
  V_{\mathrm{loc}} = \sum_{i,I}\Biggl[
    &{-}\frac{Z_{\mathrm{ion},I}}{d_{iI}}\,
      \mathrm{erf}\!\Bigl(\frac{d_{iI}}{\sqrt{2}\,r_{\mathrm{loc},I}}\Bigr)
  \\
    &{}+ e^{-\frac{1}{2}(d_{iI}/r_{\mathrm{loc},I})^2}
      \sum_{k=1}^{4} C_{k,I}
      \Bigl(\frac{d_{iI}}{r_{\mathrm{loc},I}}\Bigr)^{\!2(k-1)}
  \Biggr],
\end{split}
\end{equation}
where $Z_{\mathrm{ion},I}$ is the ionic core charge.
This matches Eq.~(1) of~\cite{hartwigsenRelativisticSeparableDualspace1998}.
The species-dependent parameters
$Z_{\mathrm{ion},I},r_{\mathrm{loc},I},C_{k,I},r_{l,I}$ and the diagonal nonlocal
couplings $h_{kk}^l,k_{kk}^l$ are tabulated for each element and indexed by the
atomic number $Z_I$.  We reconstruct the off-diagonal couplings $h_{kk'}^l,k_{kk'}^l$
($k\neq k'$) from the diagonal ones via the HGH relations.
At zero separation ($d_{iI}\to 0$) the $\mathrm{erf}/d$ term tends to
$-Z_{\mathrm{ion},I}\sqrt{2/\pi}\,/\,r_{\mathrm{loc},I}$, so $V_{\mathrm{loc}}$
is finite there by continuous extension, with coincidence value
$C_{1,I} - Z_{\mathrm{ion},I}\sqrt{2/\pi}\,/\,r_{\mathrm{loc},I}$ for each pair.  This
finiteness removes the electron--nucleus Coulomb
singularity.

The nonlocal separable part, including spin--orbit coupling, is
\begin{equation}\label{eq:vnl}
\begin{split}
  \hat V_{\mathrm{nl}} = \sum_{i,I}\sum_{l,m}\sum_{k,k'}
    &\lvert p_{k,I}^l\rangle
    \bigl[h_{kk'}^l(Z_I)
      + k_{kk'}^l(Z_I)\,L_i\!\cdot\!S_i\bigr]
  \\
    &\times\langle p_{k',I}^l\rvert,
\end{split}
\end{equation}
with $\langle\mathbf r_i|p_{k,I}^l\rangle = p_{k,I}^l(d_{iI})\,Y_{lm}$,
where the radial projectors are Gaussian and $\Gamma$-normalized functions of
$d_{iI}$.
The scalar couplings $h_{kk'}^l$ and the spin--orbit couplings $k_{kk'}^l$
define the relativistic $L_i\!\cdot\! S_i$
term.
We represent the angular shells on an auxiliary $(l,m)$ register, where
each $l$-channel projector is exact by construction.  Any deviation from the
continuum $Y_{lm}$ appears only through the grid and finite-cell discretization.
This is the operator that the block encoding implements, with the nonlocal term obtained from an exact diagonalization in the projectors' overlap metric.

The HGH pseudopotential reproduces all-electron eigenvalues to below
$10^{-5}$~a.u.\ for occupied states and $10^{-3}$~a.u.\ for unoccupied states~\cite{hartwigsenRelativisticSeparableDualspace1998},
with transferable bond
lengths.
Because core electrons are incorporated into the pseudopotential, the resulting
pseudo-wavefunctions are smoother and require a smaller dual
basis.
The pseudopotential is generated from a fully relativistic (two-component Dirac)
atomic calculation and retains relativistic effects to second order through the
spin--orbit term.  Pairing it with the non-relativistic kinetic energy is
therefore consistent (relativistic kinematics appears only via the
pseudopotential), and a scalar-relativistic variant follows by dropping the
$L_i\!\cdot\!S_i$ term.

We record several properties of HGH relevant to circuit implementation.
The pseudopotential is analytic and rapidly convergent in both real and
reciprocal space (the dual-space Gaussian form), well suited to the block
encoding.
The separable (Kleinman--Bylander) structure reduces three-center integrals to
two-center ones~\cite{kleinmanEfficaciousFormModel1982}.
Parameters are tabulated for every element from H to
Rn, so every composition
$\mathbf Z$ in the search has defined parameters.  Together, these properties imply
that every term of $H(x)$ has a closed-form dependence on the coordinates,
with species-dependent coefficients defined by the HGH parameter table indexed
by $\mathbf Z$.

Because the register is configuration-independent and, as just established, every term of $H(x)$ depends in closed form on the coordinates, we evaluate the Hamiltonian
coherently from the configuration state.  We realize this coherent evaluation as
a block encoding.

\section{Block Encoding}\label{app:block-encoding}

We establish that the Hamiltonian $H$ has a valid block encoding as a first-quantized
linear combination of unitaries
(LCU)~\cite{suFaultTolerantQuantumSimulations2021},
coherently controlled by the configuration state.  The
main result is a block-encoding theorem supported
by two lemmas: one for the HGH nonlocal pseudopotential and
one for the flag-based variable-electron embedding.  We extend the exact-subroutine result in a note after the theorem to finite arithmetic precision.

We keep three objects distinct:
(i)~the bulk material.
(ii)~$H(x)$, the discrete finite-cell Hamiltonian
  (grid truncation, minimum-image electrostatics, HGH pseudopotential), whose
  accuracy for the bulk is controlled by the approximations just listed.
(iii)~$\widetilde H(x)$, the \emph{implemented} LCU operator, in which the
  bounded potentials $V_{ee}$, $V_{nn}$, and $V_{\mathrm{loc}}$ are replaced by
  phase-sampled approximations.
The bit-expanded kinetic term is exact, as is the HGH diagonalization for
the encoded finite-grid operator (\cref{lem:hgh}), so $\widetilde H$ differs
from $H$ only by the sampling error of the three bounded terms, which
\cref{eq:sampling-error} bounds explicitly.

Throughout this appendix we take a fixed configuration $x$ as understood.  It is
written explicitly only where the configuration dependence is the point.

\emph{Assumptions.}
The theorem is conditional on the following.
\begin{enumerate}
\renewcommand{\theenumi}{\textbf{B\arabic{enumi}}}
\item \label{as:oracle}\textbf{Exact subroutines.}
  $\mathsf{PREP}$ (the PREPARE operator) prepares the amplitudes $\sqrt{c_u/\lambda}$ exactly.
  $\mathsf{QROM}$ (quantum read-only memory) returns tabulated data (HGH parameters, diagonalized
  eigenvalues, eigen-projector coefficients) exactly. $\mathsf{Arith}$ (the real-arithmetic circuit) computes
  $\mathbf G^{-1}$, distances, potentials, and $\lambda$ in exact real
  arithmetic.  We address finite precision in the
  implementation note below the theorem.
\item \label{as:caps}\textbf{Bounds for each configuration.}
  The quantities $V_{\mathrm{ee,max}}$, $V_{\mathrm{nn,max}}$, and
  $V_{\mathrm{loc,max}}$ are computed by $\mathsf{Arith}$ for each
  configuration~$x$ and are used in $\mathsf{PREP}$ and $\lambda$.  We fix the
number of phase samples $N_{\mathrm{sample}}$ globally at the
  smallest value for which the sampled-potential error bound, maximized over the allowed configuration domain,
  stays below the allocated share of the error tolerance~$\epsilon$.  Because that
  bound grows with the system size and the potential maxima,
  $N_{\mathrm{sample}}$ scales with them, not with tolerance alone.
\item \label{as:universe}\textbf{Common LCU index set.}
  We fix the index set $\{u\}$ across all~$x$, a single
  $\mathsf{PREP}/\mathsf{SEL}$ index set.  Terms that a particular
  configuration lacks have $c_u = 0$.  The electron indices in~$u$ cover all
  $N_e$ electrons and pairs.  A term with an inactive electron keeps its
  coefficient and encodes the zero operator through the flag routing.
\item \label{as:register}\textbf{Fixed-$N_e$ flagged register.}
  The register holds a fixed number $N_e$ of electrons, each extended by an
  activity flag.  For configuration~$x$, exactly
  $\sum_I Z_{\mathrm{ion},I}$ flags are set to~1, marking active physical orbitals,
  while the remaining $N_e - \sum_I Z_{\mathrm{ion},I}$ are inactive (flag~0),
  assigned to distinct padding orbitals.  We prepare antisymmetry over the
composite label (orbital $\otimes$ flag) once and preserve it throughout.
\end{enumerate}

\begin{figure*}[t]
\centering
\begin{subfigure}[c]{0.31\textwidth}
\centering
\includegraphics[width=\linewidth]{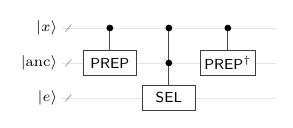}
\caption{}\label{fig:blkenc-a}
\end{subfigure}
\hfill
\begin{subfigure}[c]{0.66\textwidth}
\centering
\includegraphics[width=\linewidth]{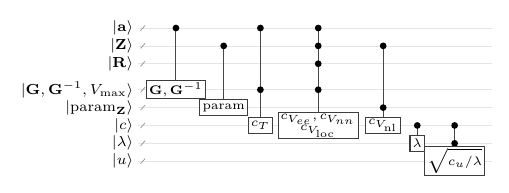}
\caption{}\label{fig:blkenc-b}
\end{subfigure}

\vspace{0.6em}
\begin{subfigure}{\textwidth}
\centering
\includegraphics[width=\linewidth]{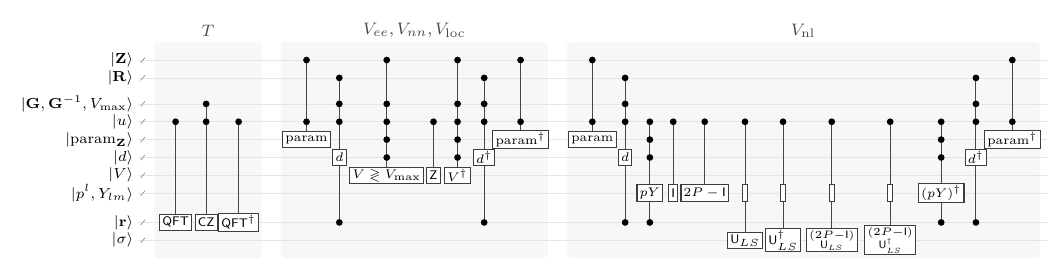}
\caption{}\label{fig:blkenc-c}
\end{subfigure}
\caption{The block-encoding circuit.  The LCU block encoding
  $\mathsf U_H = \mathsf{PREP}^\dagger\,\mathsf{SEL}\,\mathsf{PREP}$ of the
  implemented Hamiltonian $\widetilde H(x)/\lambda(x)$, for a fixed
  configuration $x$.  (a)~Abstract circuit: $\mathsf{PREP}$, $\mathsf{SEL}$, and
  $\mathsf{PREP}^\dagger$ act on the ancilla registers $\lvert\mathrm{anc}\rangle$
  and the electronic registers $\lvert e\rangle$, conditioned on the
  configuration registers $\lvert x\rangle$.  Panels (b) and (c) expand
  $\mathsf{PREP}$ and $\mathsf{SEL}$.  (b)~$\mathsf{PREP}$: arithmetic computes
  the metric tensor $\mathbf G,\mathbf G^{-1}$ from the cell structure
  $\lvert\mathbf a\rangle$.  A table lookup returns the HGH parameters for the
  species $\lvert\mathbf Z\rangle$.  Further arithmetic calculates the term
  coefficients ($c_T$; $c_{V_{ee}}, c_{V_{nn}},
  c_{V_{\mathrm{loc}}}$; $c_{V_{\mathrm{nl}}}$), sums them into the
  normalization $\lambda$, and rotates the index register $\lvert u\rangle$ to
  the amplitudes $\sqrt{c_u/\lambda}$.  (c)~$\mathsf{SEL}$ applies the unitary
  selected by the index $\lvert u\rangle$ to the electronic registers, with the
  operations grouped (shaded) by Hamiltonian term: the kinetic $T$
  ($\mathsf{QFT}$, controlled phase $\mathsf{CZ}$, $\mathsf{QFT}^\dagger$); the
  sampled bounded potentials $V_{ee}, V_{nn}, V_{\mathrm{loc}}$ (compute the
  distance $d$, compare against $V_{\max}$, apply the phase, uncompute); and the
  HGH nonlocal $V_{\mathrm{nl}}$ (build the projector $pY$, apply the shell
  reflection $2P-\mathsf I$ and the spin-orbit $\mathsf U_{LS}$, uncompute).
  Slashes mark multi-qubit registers, filled circles controls, and gate boxes
  their targets.  Daggers denote inverses.}%
\label{fig:blkenc}
\end{figure*}

\emph{LCU form.}
The implemented Hamiltonian is
\begin{equation}\label{eq:lcu}
  \widetilde H = \sum_u c_u\,\mathsf U_u,
  \quad c_u \ge 0,
  \quad \lambda = \sum_u c_u,
\end{equation}
where each $\mathsf U_u$ is unitary, with signs absorbed into the $\mathsf U_u$.
The block-encoding unitary~(\cref{fig:blkenc-a}) is
\begin{equation}\label{eq:block-enc-unitary}
  \mathsf U_H
    = \mathsf{PREP}^\dagger\,\mathsf{SEL}\,\mathsf{PREP},
\end{equation}
with state-preparation~(\cref{fig:blkenc-b}) and selection~(\cref{fig:blkenc-c}) subroutines defined by
\begin{align}
  \mathsf{PREP}\lvert 0\rangle_u
    &= \sum_u \sqrt{c_u/\lambda}\;\lvert u\rangle,
    \label{eq:prep}\\
  \mathsf{SEL}\lvert u\rangle\lvert e\rangle
    &= \lvert u\rangle\,\mathsf U_u\lvert e\rangle.
    \label{eq:sel}
\end{align}

% ---- Theorem ---------------------------------------------------------------
\begin{theorem}[Block-encoding validity (exact subroutines)]\label{thm:block-encoding}
Under assumptions \ref{as:oracle}--\ref{as:register}, the unitary
$\mathsf U_H(x)$ is an exact (zero-error)
block encoding of $\widetilde H(x)$ with normalization $\lambda(x)$.  Its
block-encoding ancilla consists of the $\lvert u\rangle$ index register, of
$\lceil\log_2 N_u\rceil$ qubits with $N_u$ the total number of LCU terms,
plus one flag-routing qubit:
\begin{equation}\label{eq:be-exact}
  (\langle 0\rvert_u \otimes I)\,\mathsf U_H(x)\,
  (\lvert 0\rangle_u \otimes I)
  = \frac{\widetilde H(x)}{\lambda(x)}.
\end{equation}
The sampling error satisfies
\begin{equation}\label{eq:sampling-error}
\begin{split}
  \bigl\lVert \widetilde H(x) - H(x) \bigr\rVert
  &\le \frac{1}{N_{\mathrm{sample}}}
  \biggl[
    \sum_{i<j} V_{\mathrm{ee,max}}(x) \\
  &\quad
    + \sum_{I<J} Z_{\mathrm{ion},I} Z_{\mathrm{ion},J}
      V_{\mathrm{nn,max}}(x)  \\
  &\quad
    + 2\Bigl(\textstyle\sum_I Z_{\mathrm{ion},I}\Bigr)
      V_{\mathrm{loc,max}}(x)
  \biggr],
\end{split}
\end{equation}
where the bracketed sum runs over the active branches, the only source of
sampling error, with the bounds computed for each configuration.
\end{theorem}

\begin{proof}
We reduce the proof to verifying that every term of $\widetilde H$ is a
finite linear combination of unitaries with nonneg\-ative coefficients.  The LCU
lemma (Lemma~1 of~\cite{suFaultTolerantQuantumSimulations2021}) then guarantees
\cref{eq:be-exact}.  We use the following unitary decompositions.

\emph{Kinetic} ($T$).  The QFT-conjugated controlled-phase construction is
  exact.

\emph{Sampled bounded potentials} ($V_{ee}$, $V_{nn}$, $V_{\mathrm{loc}}$).
  Each term is a sum of $N_{\mathrm{sample}}$ phase
  unitaries with equal coefficients.  Each term's sampling error is bounded
  by $V_{\mathrm{max}}/N_{\mathrm{sample}}$.
  Summing over terms with their multiplicities gives \cref{eq:sampling-error}.

\emph{Nonlocal pseudopotential} ($\hat V_{\mathrm{nl}}$).  \Cref{lem:hgh}
  provides a finite unitary LCU with nonneg\-ative coefficients.

The three decompositions cover every term of $\widetilde H$, and
\cref{lem:flag} shows that our flag-controlled construction preserves the
unitarity of each $\mathsf U_u$ and the LCU structure on the fixed-$N_e$
register.  Because all arithmetic, QROM lookups, flag controls, and selectors are
controlled by the untouched configuration register, $\mathsf U_H$ is
block-diagonal in~$x$ and constitutes our \emph{variable-normalization block
encoding} of $\bigoplus_x \widetilde H(x)/\lambda(x)$.  The
normalization $\lambda(x)$ is computed coherently into a register.  The common index set ensures that dependence on $x$
appears only through the coefficients $c_u(x)$ (zero for absent
terms) and the data-controlled unitaries.  A standard
single-normalization encoding follows by rescaling every configuration to
$\lambda_{\max}$, giving each the same uniform normalization.
\end{proof}

\emph{Implementation note (finite precision).}
Under finite $b$-bit $\mathsf{Arith}$ and finite-precision amplitude preparation
we obtain an approximate block encoding with the same
normalization $\lambda$, a bounded error $\epsilon$, and an enlarged
ancilla register (the index register plus arithmetic ancillas).
The error $\epsilon$ decreases with the bit precision $b$.

% ---- Channel decomposition -------------------------------------------------
\subsection{Term decomposition and the normalization $\lambda$}%
\label{sec:be-channels}

\subsubsection{Kinetic energy}\label{sec:be-kinetic}

Each momentum vector $\mathbf k_i$ is stored in signed binary.  Its $t$-th-bit
slice $\mathbf k_{i,t}$ has entries in $\{-1,0,1\}$, one per Cartesian axis.
Each entry is a magnitude bit ($0$ or $1$) signed by that axis's sign bit.  Expanding the quadratic form
(\cref{eq:kinetic}) over bit positions gives the exact identity
\begin{equation}\label{eq:kinetic-bitslice}
  T = 2\pi^2 \sum_{i\;\mathrm{active}} \sum_{t,t'} 2^{t+t'}\,
      \mathbf k_{i,t}^{\,\top}\,\mathbf G^{-1}(\mathbf a)\,
      \mathbf k_{i,t'}
\end{equation}
(reciprocal-metric quadratic form for the general
cell).
Following the bit-slice linearization (Eq.~(23) of~\cite{suFaultTolerantQuantumSimulations2021}),
each scalar bit product $m_{t}\,m_{t'} = \tfrac{1}{2}(1 - (-1)^{m_{t}\,m_{t'}})$
produces \emph{two} LCU unitaries per bit pair (an identity part and a
controlled-phase part), with the phase unitary
$\mathsf U_u = \mathsf{QFT}^\dagger\,\mathsf{CZ}_u\,\mathsf{QFT}$.
Each $\mathsf{CZ}_u$ gate has the coefficient
$\pi^2\,2^{t+t'}\,\lvert(\mathbf G^{-1})_{xy}\rvert$, with $x,y$ the Cartesian
axes of bits $t,t'$, so the combined identity--phase pair contributes twice that.

We note two features of the contraction.
(a)~For a squared-component term ($x{=}y$, giving $(\mathbf G^{-1})_{xx}$),
the sign factor squares to $1$ and the first (``identity'')
part is $\mathsf I$.
For a cross-component term ($x{\neq}y$, giving $(\mathbf G^{-1})_{xy}$), however, the first
part is the product of the two axes' sign bits, a
$\mathsf Z \otimes \mathsf Z$ on the two sign qubits, \emph{not} the literal
$\mathsf I$.  Only the second part adds the $\mathsf{CZ}$ on the magnitude
bits.  We absorb both the sign of $(\mathbf G^{-1})_{xy}$ and the minus sign
from the phase part into $\mathsf{CZ}_u$.
(b)~The contraction runs over all ordered axis pairs.  Each off-diagonal
$(\mathbf G^{-1})_{xy}$ appears at its two symmetric
positions ($xy$ and $yx$), each counted once.  No additional factor of two multiplies the
off-diagonal coefficient.  Such a factor would double-count.
Together, the identity and phase unitaries encode $T$ exactly.  There is no
omitted constant offset.

The kinetic contribution to the 1-norm is
\begin{equation}\label{eq:lambda-T}
  \lambda_T = 2\pi^2 \sum_{i=1}^{N_e}
  \sum_{t,t'} \sum_{x,y} 2^{t+t'}\,
  \lvert(\mathbf G^{-1})_{xy}\rvert.
\end{equation}

\subsubsection{Sampled bounded potentials}\label{sec:be-sampled}

We sample each bounded term ($\lvert V\rvert \le V_{\mathrm{max}}$,
nonneg\-ative kernel) into $N_{\mathrm{sample}}$ phase unitaries with equal
coefficients and error at most $V_{\mathrm{max}}/N_{\mathrm{sample}}$ in
each term.
Three terms use this mechanism.  In each case the
bound is computed for each configuration by $\mathsf{Arith}$ and appears in
$\mathsf{PREP}$ and $\lambda$.

\emph{Electron--electron repulsion} ($V_{ee}$).
The kernel $1/d_{ij}$ is positive.  The maximum resolution limit is
$V_{\mathrm{ee,max}} = 1/d_{\min}(\mathbf a)$, the inverse of the shortest
distinct grid-point separation under the metric of \cref{eq:min-image}.
The unordered-pair convention ($i<j$) absorbs $1/d$
directly.  The
multiplicity is the number of unordered pairs $i < j$.  This gives
$\lambda_{V_{ee}} = \binom{N_e}{2}\,
V_{\mathrm{ee,max}}$.

\emph{Nucleus--nucleus repulsion} ($V_{nn}$).
The kernel $Z_{\mathrm{ion},I} Z_{\mathrm{ion},J}/d_{IJ}$ is positive.  The
bound for each pair is $Z_{\mathrm{ion},I} Z_{\mathrm{ion},J} V_{\mathrm{nn,max}}$,
with multiplicity over pairs $I < J$.  This operator is diagonal in the
nuclear position and species registers ($\propto \mathsf I$ on the electronic
register), but we include it \emph{inside} the block encoding. $V_{nn}$ shifts
every eigenvalue of $H$ and is captured in the qubitized
$\widetilde E_n$.  The contribution is
$\lambda_{V_{nn}} = \sum_{I<J} Z_{\mathrm{ion},I} Z_{\mathrm{ion},J}
V_{\mathrm{nn,max}}$.

\emph{Local pseudopotential} ($V_{\mathrm{loc}}$).
The kernel (\cref{eq:vloc}) is indefinite.  The $-Z_{\mathrm{ion}}
\mathrm{erf}/d$ term is attractive, while the Gaussian-polynomial terms can be
positive.  Since phase sampling requires a nonneg\-ative
kernel, we split
$V_{\mathrm{loc}} = V_+ - V_-$ with
$V_\pm = \max(\pm V_{\mathrm{loc}}, 0) \ge 0$ and sample each part
separately, absorbing the minus sign of $V_-$ into its unitaries.
The bound $V_{\mathrm{loc,max}}$ is the \emph{maximum} of
$\lvert V_{\mathrm{loc}}\rvert$ over the grid, bounding both signed
parts $V_+$ and $V_-$.  It is computed for each configuration by
$\mathsf{Arith}$.  The origin-value formula
$\sum_I \lvert C_{1,I} - Z_{\mathrm{ion},I}
r_{\mathrm{loc},I}^{-1}\sqrt{2/\pi}\rvert$
gives $\lvert V_{\mathrm{loc}}\rvert$ at $d_{iI} = 0$.  Because the $C_2$--$C_4$
Gaussian-polynomial terms need not be largest at $d_{iI} = 0$, the grid
maximum $V_{\mathrm{loc,max}}$ may exceed it.
The nucleus sum is performed inside the computed potential.  Sampling proceeds
over each electron~$i$ (multiplicity $N_e$) and over
the two parts $V_+$, $V_-$, giving
$\lambda_{V_{\mathrm{loc}}} = 2N_e\,
V_{\mathrm{loc,max}}$.

We determine $N_{\mathrm{sample}}$ from the global maxima of the three bounds over the configuration domain, so that each term's sampling error
$V_{\mathrm{max}}/N_{\mathrm{sample}}$ meets the uniform tolerance target
for every~$x$.

\subsubsection{LCU term summary}\label{sec:be-table}

\Cref{tab:lcu-channels} collects the LCU index, coefficient,
and $\lambda$-contribution of every term.  The total normalization is
$\lambda = \sum_u \lvert c_u\rvert$, with electron indices covering all
$N_e$ electrons.  Inactive electrons contribute to $\lambda$ alone.  No $m$ multiplicity appears in the nonlocal term
because we treat the full $l$-shell as a single reflection.  The factor of two
in $\lambda_{V_{\mathrm{loc}}}$ arises
from the common $V_\pm$ bound.

\begin{table*}[t]
\caption{LCU term decomposition.  Each row lists the LCU index $u$
  (the quantum numbers selecting a unitary), the coefficient $c_u$,
  and the contribution to the 1-norm $\lambda$.  Bit positions $t,t'$ and
  Cartesian axes $x,y$ index the momentum register.
  $m\le N_{\mathrm{sample}}$ labels the phase sample.}\label{tab:lcu-channels}
\begin{ruledtabular}
\begin{tabular}{llcc}
Term
  & $u$
  & $c_u$
  & $\lambda$-contribution \\
\colrule
$T$
  & $(i;\,t,t';\,xy)$, identity/phase
  & $\pi^2\,2^{t+t'}\,\lvert(\mathbf G^{-1})_{xy}\rvert$
  & $2\pi^2\!\sum_i \sum_{t,t'}\sum_{x,y} 2^{t+t'}\,\lvert(\mathbf G^{-1})_{xy}\rvert$
  \\
$V_{ee}$
  & $(i{<}j,\; m)$
  & $V_{\mathrm{ee,max}}/N_{\mathrm{sample}}$
  & $\binom{N_e}{2}\,V_{\mathrm{ee,max}}$
  \\
$V_{nn}$
  & $(I{<}J,\; m)$
  & $Z_{\mathrm{ion},I} Z_{\mathrm{ion},J}\,
    V_{\mathrm{nn,max}}/N_{\mathrm{sample}}$
  & $\sum_{I<J} Z_{\mathrm{ion},I} Z_{\mathrm{ion},J}\,
    V_{\mathrm{nn,max}}$
  \\
$V_{\mathrm{loc}}$
  & $(i,\;\pm,\; m)$
  & $V_{\mathrm{loc,max}}/N_{\mathrm{sample}}$
  & $2N_e\,V_{\mathrm{loc,max}}$
  \\
$\hat V_{\mathrm{nl}}$ (scalar)
  & $(i,\,I,\,l,\,k,\;\{\mathsf I,\,2P{-}\mathsf I\})$
  & $\lvert h(Z)\rvert/2$
  & $\sum_{i,I,l,k}\lvert h(Z)\rvert$
  \\
$\hat V_{\mathrm{nl}}$ (spin--orbit)
  & $(i,\,I,\,l{>}0,\,k,\;\{1,\dots,4\})$
  & $\lvert k(Z)\rvert(l+1)/8$
  & $\sum_{i,I,l>0,k}\lvert k(Z)\rvert(l+1)/2$
  \\
\end{tabular}
\end{ruledtabular}
\end{table*}

\subsection{HGH nonlocal LCU}\label{sec:be-nonlocal}

\begin{lemma}[HGH nonlocal: overlap-metric diagonalization and full
$l$-shell spin--orbit LCU]\label{lem:hgh}
The HGH nonlocal operator $\hat V_{\mathrm{nl}}$ of \cref{eq:vnl} has a
finite unitary LCU with nonneg\-ative coefficients, implementable given the
subroutines stated below.  A first-quantized block encoding of the
\emph{scalar} separable projector was given in Eq.~(13)
of~\cite{berryQuantumSimulationRealistic2024}.  We introduce the
overlap-metric diagonalization and the spin--orbit ($l>0$) channel here.
For
each electron~$i$, nucleus~$I$, angular momentum~$l$, and radial
eigen-index~$k$, the LCU contains:
\begin{enumerate}
\item \textbf{Scalar channel.}  Two unitaries
  $\mathsf U_{u,h} \in
  \bigl\{\mathrm{sgn}(h)\,\mathsf I,\;
         \mathrm{sgn}(h)\,(2P - \mathsf I)\bigr\}$
  with coefficient $c_{u,h} = \lvert h(Z)\rvert/2$.
\item \textbf{Spin--orbit channel} ($l > 0$).  Four unitaries
  $\mathsf U_{u,k} \in
  \bigl\{\mathrm{sgn}(k)\,\mathsf U_{LS},\;
         \mathrm{sgn}(k)\,\mathsf U_{LS}^\dagger,\;
         \mathrm{sgn}(k)\,(2P-\mathsf I)\,\mathsf U_{LS},\;
         \mathrm{sgn}(k)\,(2P-\mathsf I)\,\mathsf U_{LS}^\dagger\bigr\}$
  with coefficient $c_{u,k} = \lvert k(Z)\rvert(l+1)/8$.
\end{enumerate}
Here
$P = \lvert p_k^l\rangle\langle p_k^l\rvert \otimes \Pi_l
     \otimes \mathsf I_{\mathrm{spin}}$
is the rank-$2(2l+1)$ projector formed from the radial eigen-projector, the
angular $l$-shell projector
$\Pi_l = \sum_m \lvert Y_{lm}\rangle\langle Y_{lm}\rvert$, and the identity on
the spin register, so that $[P,\,L_i\!\cdot\!S_i] = 0$.  The unitary
$\mathsf U_{LS} = A_l + i\sqrt{\mathsf I - A_l^2}$ is constructed from
$A_l = \tfrac{2}{l+1}\,\Pi_l(L_i\!\cdot\!S_i)\Pi_l$, which acts on the
$2(2l+1)$-dimensional angular-$\otimes$-spin shell.
\end{lemma}

\begin{proof}
We proceed in three steps: diagonalization of the radial coupling
matrices, construction of the full $l$-shell projector, and assembly of the
unitary LCU.

\emph{Step~1: Overlap-metric diagonalization.}
For a fixed species $Z_I$ and angular momentum~$l$, the scalar nonlocal
operator on electron~$i$ is
$\sum_m \sum_{kk'} \lvert p_k^l Y_{lm}\rangle\,h_{kk'}^l\,
\langle p_{k'}^l Y_{lm}\rvert$,
where
$h^l$ is a real symmetric matrix.  The radial projectors $p_k^l$ are
unit-normalized in the continuum (Eq.~(4) of~\cite{hartwigsenRelativisticSeparableDualspace1998})
but are
\emph{not} mutually orthogonal.  Equal-width Gaussians centered at the same
nucleus overlap.  We therefore obtain the physically correct spectrum by
diagonalizing $h^l$ in the overlap metric
$\langle p_k^l \vert p_{k'}^l\rangle$, a Hermitian generalized eigenproblem.  Its
eigenvalues $h(Z)$ are the scalar couplings of each term.  The corresponding
eigenvectors yield mutually orthonormal radial eigen-projectors
$\lvert p_k^l\rangle$.  We diagonalize the
spin--orbit matrix $k^l$ separately in the same overlap metric, producing
its own
eigenvalues $k(Z)$ and its own set of orthonormal eigen-projectors.  The scalar
and spin--orbit channels do not in general share eigenvectors, so we store
separate precomputed data for them in the QROM.

\emph{Auxiliary angular-channel decomposition.}
We implement the nonlocal channel not by discretizing the spherical
harmonics $Y_{lm}$ on the real-space grid (which would not yield exactly
orthonormal angular functions) but by keeping the angular labels $(l,m)$
explicitly on an auxiliary register.  On this register
$\Pi_l = \sum_m \lvert Y_{lm}\rangle\langle Y_{lm}\rvert$ is by construction
the exact orthogonal projector onto the $l$-shell, hence a true projector and
$L_i\!\cdot\!S_i$-invariant ($[\Pi_l,\,L_i\!\cdot\!S_i] = 0$).  We apply the
Kleinman--Bylander projectors through this register from the analytic HGH
dual-space form factors (the preparation
$\mathsf U_{\mathsf{PREP}}$)~\cite{berryQuantumSimulationRealistic2024}.
The radial overlap-metric diagonalization is therefore \emph{exact for the
implemented operator}.  The difference between the continuum HGH nonlocal
operator and this finite auxiliary construction belongs to the grid
approximation of the Hamiltonian (the $H$-versus-bulk difference), not to an
$\widetilde H$-versus-$H$ error, consistent with the claim at the opening of this
appendix.  Computing the full $(k,m)$
grid overlaps would instead mix $m$ values and complicate the shell
structure.

\emph{Step~2: Full $l$-shell projector.}
Define
$P = \lvert p_k^l\rangle\langle p_k^l\rvert \otimes \Pi_l
     \otimes \mathsf I_{\mathrm{spin}}$,
a rank-$2(2l+1)$ projector combining the radial eigen-projector with the full
angular $l$-shell
$\Pi_l = \sum_m \lvert Y_{lm}\rangle\langle Y_{lm}\rvert$ and the identity on
the two-dimensional spin space.
The reflection is $2P - \mathsf I$.  Because $P$ contains
$\Pi_l \otimes \mathsf I_{\mathrm{spin}}$, the commutator
$[P,\,L_i\!\cdot\!S_i] = 0$ holds. $\Pi_l$ projects onto the $l$-shell, an
eigenspace of $L^2$ and hence an $L_i\!\cdot\!S_i$-invariant subspace, while
$\mathsf I_{\mathrm{spin}}$ is the identity on spin and the radial
eigen-projector acts on the untouched radial factor.

For the \emph{scalar} channel, the identity
$h(Z)\,P = \tfrac{h(Z)}{2}(\mathsf I + (2P - \mathsf I))$
splits one eigenvalue term into two unitaries (identity and reflection), each
with coefficient $\lvert h(Z)\rvert/2$ and the sign of $h(Z)$ absorbed into the
unitary.

For the \emph{spin--orbit} channel ($l > 0$), define the restricted angular
momentum operator
$A_l = \tfrac{2}{l+1}\,\Pi_l (L_i\!\cdot\!S_i)\Pi_l$,
a Hermitian operator on the $2(2l+1)$-dimensional angular-$\otimes$-spin shell.
Within the $l$-shell, $L_i\!\cdot\!S_i$ has eigenvalue $l/2$ at total angular
momentum $j = l + \tfrac{1}{2}$ and eigenvalue $-(l+1)/2$ at
$j = l - \tfrac{1}{2}$, so the eigenvalues of $A_l$ are $l/(l+1)$ and $-1$,
giving $\lVert A_l\rVert \le 1$.  Therefore
$\mathsf U_{LS} = A_l + i\sqrt{\mathsf I - A_l^2}$
is unitary on the $2(2l+1)$-dimensional space, and
$A_l = \tfrac{1}{2}(\mathsf U_{LS} + \mathsf U_{LS}^\dagger)$.
Since $P$ commutes with $L_i\!\cdot\!S_i$ (and hence with $A_l$), we have
$P(L_i\!\cdot\!S_i) = \tfrac{l+1}{2}\,P A_l$.  This product is Hermitian.  The
spin--orbit eigenvalue term then decomposes as
\begin{equation}\label{eq:so-lcu}
  k(Z)\,P(L_i\!\cdot\!S_i)
  = \frac{k(Z)(l+1)}{8}\,
    (\mathsf I + (2P-\mathsf I))
    (\mathsf U_{LS} + \mathsf U_{LS}^\dagger),
\end{equation}
a product of two two-term sums yielding four unitaries, each with coefficient
$\lvert k(Z)\rvert(l+1)/8$.

A rank-one projector for a single $m$,
$\lvert p_k^l Y_{lm}\rangle\langle p_k^l Y_{lm}\rvert$, would \emph{not}
commute with $L_i\!\cdot\!S_i$ and would increase $\lambda$ by a factor of
$2l+1$ through internal cancellation.  The full $l$-shell projector avoids this
by absorbing the $m$ multiplicity into a single reflection.

\emph{Step~3: Unitarity, coefficients, and reflection construction.}
Each listed $\mathsf U_{u,h}$ and $\mathsf U_{u,k}$ is unitary
($\mathsf I$, a reflection, $\mathsf U_{LS}$, or products thereof).  All
coefficients are nonneg\-ative after taking absolute values, with signs in the
unitaries.  The identity--reflection split preserves the induced 1-norm.

The reflection $(2P - \mathsf I)$ is realized through a state-preparation
operator $\mathsf U_{\mathsf{PREP}}$.  Working in the nucleus-relative
coordinate (a shift $\mathbf r \to \mathbf r - \mathbf R_I$ performed by
$\mathsf{Arith}$), this operator prepares the shell projector wavefunctions
from the analytic HGH dual-space Gaussian form
factors, with parameters
loaded from QROM.  The operator acts as a unitary
basis change on $e \otimes m \otimes \sigma$, in which the shell label $m$
moves from the $m$-register into the prepared wavefunction on the electron
register, the $m$-register returns to $\lvert 0\rangle$, and spin is
untouched:
\begin{equation}\label{eq:uprep}
  \mathsf U_{\mathsf{PREP}}\,\lvert 0\rangle_e\,
  \lvert m\rangle\,\lvert\sigma\rangle
  = \lvert p_k^l\,Y_{lm}\rangle_e\,
  \lvert 0\rangle_m\,\lvert\sigma\rangle,
  \quad m = -l,\dots,l.
\end{equation}
The map is an isometry on the $(2l{+}1)$-dimensional shell subspace,
and it extends to a unitary basis change on the joint space.
The reflection is then \emph{exact} given this subroutine:
\begin{equation}\label{eq:reflection}
  2P - \mathsf I
  = \mathsf U_{\mathsf{PREP}}\,
    \bigl(2\lvert 0\rangle\langle 0\rvert_e
          \otimes \mathsf I_{m,\sigma}
          - \mathsf I\bigr)\,
    \mathsf U_{\mathsf{PREP}}^\dagger.
\end{equation}
The flag is ``electron register $= \lvert 0\rangle_e$''.  The inner
reflection records whether the electron register is in the trial state.
The internal shell degrees of freedom, tracked through the conjugated
basis by the $m$-register and the spin register, are
\emph{preserved}.
The reflection marks a flag on the $P$-subspace but does not collapse the
rank-$2(2l+1)$ shell.

\emph{Register matching.}
$P$ acts on $e\otimes\sigma$, while $\mathsf U_{\mathsf{PREP}}$ acts on
$e\otimes m\otimes\sigma$.
Since $\mathsf U_{\mathsf{PREP}}$ clears the $m$-register to its reference
$\lvert 0\rangle$ (the label $m$ is uncomputed into the prepared
wavefunction), the reflection equals the operator
$2(P\otimes\lvert 0\rangle\langle 0\rvert_m)
- \mathsf I$.  With the $m$-register initialized to $\lvert 0\rangle$ and
restored after every application, this operator implements $2P - \mathsf I$ on the
physical electron--spin register $e\otimes\sigma$.
$\mathsf U_{\mathsf{PREP}}$ is the Kleinman--Bylander projector preparation,
realizable from the analytic form factors.

We take the unitary $\mathsf U_{LS}$ as a fixed $2(2l+1) \times 2(2l+1)$ matrix on
the angular-$\otimes$-spin shell, built once for each (species,~$l$).
For each species, we precompute the diagonalized data from the HGH parameter
table (Table~I of~\cite{hartwigsenRelativisticSeparableDualspace1998}):
eigenvalues $h(Z)$, $k(Z)$ and orthonormal radial eigenfunction coefficients,
separate for scalar and spin--orbit channels, each
$h^l$ and $k^l$ assembled from its tabulated diagonal and the reconstructed
off-diagonal couplings.  These eigendecompositions replace the matrices $h_{kk'}^l$,
$k_{kk'}^l$ in the QROM.
\end{proof}

\subsection{Flag-based active-electron embedding}\label{sec:be-flag}

\begin{lemma}[Flag-embedding and configuration lift]\label{lem:flag}
There is an embedding of the physical
$\sum_I Z_{\mathrm{ion},I}$-electron antisymmetric states into the
fixed-$N_e$ flagged register such that, restricted to the image of this
embedding, $\widetilde H(x)$ acts as the physical active-electron
Hamiltonian.
The image of the embedding is invariant under
$\widetilde H(x)$, and the initial state lies in this image.
The resulting $\mathsf U_H$ is block-diagonal in~$x$ with
normalization $\lambda(x)$.
\end{lemma}

\begin{proof}
\emph{Construction.}
We extend each single-particle label by one qubit: active (physical
orbital, flag~1) or inactive (distinct padding orbital, flag~0).  The
configuration~$x$ sets exactly $\sum_I Z_{\mathrm{ion},I}$ flags to~1.  The
embedding sends a physical antisymmetric state of
$\sum_I Z_{\mathrm{ion},I}$ active electrons to its antisymmetrized
product with a fixed set of $N_e - \sum_I Z_{\mathrm{ion},I}$ inactive padding
orbitals, distinct to enforce Pauli exclusion.  The embedding is isometric.  Every
term of $\widetilde H$ on the fixed-$N_e$ flagged register is
flag-controlled.  One-body operators $T$, $V_{\mathrm{loc}}$,
$\hat V_{\mathrm{nl}}$ act on electron~$i$ only if its flag is~1.  The two-body
operator $V_{ee}$ acts on a pair of electrons only if both their flags are~1.  No term flips a flag.
Flag control acts by routing.  For each term that acts on electron~$i$,
$\mathsf{SEL}$ flips a routing qubit, part of the block-encoding ancilla,
when the flag of~$i$ is~0 (for $V_{ee}$, when either flag is~0).  Branches
with the routing qubit set lie outside the encoded block, so the block acts
as the zero operator on inactive branches.

\emph{Invariance and the embedding identity.}
Flag control implies that $\widetilde H$ on the flagged register commutes
with the flag-occupation number and never references padding orbitals.
Consequently, the image of the embedding is invariant under
$\widetilde H$, and restricted to this image, $\widetilde H$ equals the
physical Hamiltonian.  Antisymmetry over the composite label
(orbital $\otimes$ flag) is imposed once at state preparation and is preserved
by $[\widetilde H,\,\mathrm{perm}] = 0$.
The inactive electrons do not interact.  They contribute zero energy and occupy a
fixed antisymmetrized product of padding orbitals.  This product is
label-correlated with the active electrons---not a separate tensor
factor---yet does not couple to the active dynamics.

\emph{Restriction of dynamics.}
The prepared initial state lies in the image of the embedding, and
$\widetilde H$ preserves this image, so the encoded dynamics remain within it.
Every operation acts within the image of the embedding, or orthogonally to it,
and unphysical flag sectors do not interfere.

\emph{Configuration lift.}
We make the encoding block-diagonal in~$x$ by the configuration-controlled arithmetic, QROM lookups, flag controls, and
selectors (the configuration lift), realizing
$\bigoplus_x \widetilde H(x)/\lambda(x)$, as in \cref{thm:block-encoding}.
\end{proof}

\medskip
Under the stated assumptions, we obtain from the theorem and
the finite-precision implementation note, respectively, an exact and an
approximate block encoding of $\widetilde H$ across the configuration
superposition.

\section{State Isolation}\label{app:isolation}

The block encoding provides, for each
configuration, a unitary $\mathsf U_H(x)$ that encodes the
normalized implemented Hamiltonian $\widetilde H(x)/\lambda(x)$.  We construct a
single fixed unitary (a product of $m_d$ discrete walk operators) that
evolves every configuration's electronic state from the corresponding eigenstate of the
degeneracy-broken kinetic energy $T + \eta D$ to the target eigenstate
of the degeneracy-broken implemented Hamiltonian $\widetilde H(x) + \eta D$, coherently and simultaneously
across the configuration superposition.

The degeneracy-breaking term $\eta D$ makes the followed eigenvalue (the eigenvalue tracked along the path)
typically simple.  The target state is
$\lvert\Psi_{\mathrm{target}}\rangle$ of the main text, where $\psi_n(x)$ is the
eigenstate of $\widetilde H(x) + \eta D$.  The output
differs from $\lvert\Psi_{\mathrm{target}}\rangle$ only by a relative
phase on each configuration.  Expanding in the eigenbasis with
coefficients $c_n(x)$, one has
$\lvert c_n(x)\rvert \approx 1$ and $c_{m\neq n}(x) \approx 0$ for every
configuration.
Each configuration's phases need not be tracked here.

The isolation must be a single fixed unitary, uniform across the
configuration superposition: no measurement, no
repeat-until-success protocol tied to individual configurations, no number of amplitude-amplification iterations
that depends on each configuration's overlaps or energies.
A ground-state filter with query cost linear in the inverse gap suits a
superposition in which one energy threshold lies within every configuration's
gap, as for nearby geometries of one
molecule~\cite{simonImprovedPrecisionScaling2024}.  Across compositions the
ground energies span a range far wider than any gap, and discrete adiabatic
evolution meets the requirement for every configuration.
We apply the discrete adiabatic large-time-step
framework~\cite{anLargeTimestepDiscretisation2025}, built on the explicit-gap
discrete adiabatic theorem~\cite{costaOptimalScalingQuantum2022}.  A fixed
sequence of $m_d$ walk operators, applied to every configuration
simultaneously, evolves the state along the interpolation path with no
configuration-dependent control.

\emph{Interpolation and degeneracy breaking.}
The interpolating Hamiltonian is
$H_s = (1-f(s))\,H_0 + f(s)\,H_1$ with $H_0 = T + \eta D$ and
$H_1 = \widetilde H(x) + \eta D$, so that
\begin{equation}\label{eq:interp}
  H_s = T + \eta D + f(s)\,V,
\end{equation}
where $V = U_{en} + V_{ee} + V_{nn}$ is the potential, phase-sampled in
the implemented encoding, and the
coefficient of $D$ is the constant $\eta$, not scheduled.
Every operator in this appendix is realized by the block encoding. $H_s$ and $H_1$ are the implemented
interpolating Hamiltonians, exact in the kinetic and degeneracy-breaking terms
$T + \eta D$, and phase-sampled in the potential.  The walk's
endpoint is therefore the target eigenstate of $\widetilde H(x) + \eta D$.
Because \cref{eq:sampling-error} bounds $\lVert\widetilde H(x) - H(x)\rVert$
uniformly over the configuration domain, Weyl's inequality gives
\begin{equation}\label{eq:spectral-transfer}
  \bigl\lvert \widetilde E_n(x) - E_n(x)\bigr\rvert
  \le \bigl\lVert \widetilde H(x) - H(x)\bigr\rVert
\end{equation}
for every eigenvalue, and any spectral gap of $H(x)$ separating $E_n(x)$
from the rest of its spectrum is inherited by $\widetilde H(x)$, shrinking
by at most $2\lVert\widetilde H(x) - H(x)\rVert$.  The implemented spectrum and
gap match those of the ideal (unsampled) Hamiltonian within this bound.

We take the degeneracy-breaking one-body field $D$ to be fixed and
configuration-independent, diagonal in momentum and spin.  It
assigns a strict ordering to the single-particle plane-wave states, splitting
$\pm\mathbf k$ pairs, momentum shells, and spin.
It has unit block-encoding normalization ($\sum_{u\in D}\lvert c_u\rvert = 1$),
so its $\lambda$-contribution is exactly $\eta$.
It breaks the symmetry-forced degeneracies at both endpoints
(the open-shell ground space of $T$ at $s=0$ and symmetry-forced target
degeneracies of $H(x)$ at $s=1$), so the followed eigenvalue is
typically simple along the whole path.  A fixed one-body $D$ need not
split every many-body degeneracy.  Any remaining degeneracy is excluded
by assumption~\ref{as:gap}.

The target is the eigenstate of $\widetilde H(x) + \eta D$, a single well-defined state at
finite $\eta$.  We control the $\eta$-dependence below.
By first-order perturbation theory, provided the target eigenspace of
$H(x)$ is isolated from the non-target spectrum, the prepared state
differs from the target eigenstate of $H(x)$ by a component that vanishes
linearly in $\eta$.  The proportionality constant is set by that
isolating spectral gap.  The prepared state is not, in general, an exact
eigenstate of $H(x)$.
The degeneracy-breaking strength $\eta$ is a precomputed classical constant.  It sets the
lower gap bound at the broken degeneracies (contributing to $\Delta$) and introduces an
$\mathcal O(\eta)$ selection bias in evaluated properties.  We choose it
within the overall error allocation.

For each configuration, the initial eigenstate is the unique ground state of
$T(\mathbf a) + \eta D$, the degeneracy-broken plane-wave Slater determinant
formed by antisymmetrizing the lowest single-particle states under the strict $D$ ordering
of kinetic energies $\mathbf G^{-1}(\mathbf a)$.
This state is part of the prepared initial superposition
$\lvert\Psi_{\mathrm{init}}\rangle$ of the main text.  The
occupied plane-wave set is held fixed across configurations by the domain restriction of
assumption~\ref{as:init}, making the initial state a single configuration-independent
determinant.

\emph{Assumptions.}
The convergence bounds below are conditional on the following.
\begin{enumerate}
\renewcommand{\theenumi}{\textbf{C\arabic{enumi}}}
\item \label{as:domain}\textbf{Bounded configuration domain (uniform
  constants).}
  The allowed configuration domain is finite (discrete registers) and
  contains only non-degenerate cells (Niggli-reduced, positive volume), so $\lambda_T(x)>0$ for every configuration and
  the potential coefficient ratio
  $\max_x \sum_{u\notin T\cup D}\lvert c_u(x)\rvert / \lambda_T(x)$
  is finite.  The degeneracy-breaking field contributes only the fixed
  constant $\eta$.
  This is the same maximization over the allowed domain used to fix $N_{\mathrm{sample}}$.  It is the
  assumption under which the constants appearing in the
  $\mathcal O((\lambda/\Delta)^{3}\epsilon^{-1})$ bound of Tier~1 below are uniform
  across configurations.

\item \label{as:gap}\textbf{Uniform gap and simplicity along the path.}
  For every configuration $x$ in the allowed domain and every
  $s\in[0,1]$, the followed eigenvalue of $H_s(x)$ is simple and
  separated from the rest of the spectrum by at least $\Delta > 0$, a
  uniform physical gap.  The corresponding gap of the normalized
  $H_s/\lambda(x)$ is at least $\Delta/\lambda(x)$.
  The degeneracy breaking makes this assumption typically satisfiable. $D$ removes the
  symmetry-forced degeneracies, and the transition from the $D$ ordering to
  the physical ordering (as $f(s)\,V$ grows past $\eta D$) proceeds
  through avoided crossings whose minimum gaps are absorbed into $\Delta$.
  Accidental many-body near-degeneracies remain covered by the assumption
  itself, the standard precondition of adiabatic eigenstate preparation.
  The multistep gap condition (a gap also between successive discrete
  steps) is implied by this uniform gap once $m_d$ is large
  enough (Lemma~20 of~\cite{anLargeTimestepDiscretisation2025}).

\item \label{as:init}\textbf{Initial state.}
  For each configuration, the prepared electronic state is the unique ground state of
  $T(\mathbf a) + \eta D$, the degeneracy-broken plane-wave Slater
  determinant.
  We restrict the allowed cell domain so that the $N_e$ lowest-energy
  plane-wave states form the same set for every configuration.  A uniform margin, larger than $\eta$, separates the $N_e$-th and $(N_e{+}1)$-th kinetic-energy levels across the
  domain, so neither cell anisotropy via
  $\mathbf G^{-1}(\mathbf a)$ nor the degeneracy-breaking term $\eta D$ reorders the
  occupied set.  The initial determinant is then a single fixed
  occupation (the same plane-wave labels filled on every configuration), and
  the breaking removes the open-shell ambiguity, so no configuration-dependent
  selection input is needed.  The cost formulas assume this restricted cell
  domain.

\item \label{as:highorder}\textbf{High-order discrete adiabatic theorem
  (conditional tier only).}
  We additionally assume the high-order
  discrete adiabatic theorem for the superpolynomial-convergence tier, presently available as a conjecture.  Its
  original proof has a known missing step, and the statement is supported by numerical
  evidence (Conjecture~10 of~\cite{anLargeTimestepDiscretisation2025}).
\end{enumerate}

\emph{Schedule.}
We normalize the glue function $g$ so
that $g(0)=0$, $g(1)=1$, and $g^{(k)}(0) = g^{(k)}(1) = 0$ for every $k\ge 1$,
the boundary cancellation condition.
We take the schedule as the symmetric composite
\begin{equation}\label{eq:schedule}
  f(s) = \begin{cases}
    \tfrac{1}{2}\,g(2s), & s\in[0,\tfrac{1}{2}],\\[2pt]
    \tfrac{1}{2} + \tfrac{1}{2}\,g(2s-1), & s\in(\tfrac{1}{2},1].
  \end{cases}
\end{equation}
The symmetric composite inherits this smoothness and additionally slows in
the middle of the path, where interpolation gaps typically close.
Our schedule is fixed and problem-independent.  It requires no knowledge of
the gap $\Delta$, of the configuration $x$, or of the system size.
We discretize the path into $m_d$ steps with
$s = j/(m_d-1)$, $j\in\{0,\dots,m_d-1\}$, so the endpoints $s=0$ and $s=1$
are both included.  The values $f(j/(m_d-1))$ are
precomputed classical constants.
The degeneracy-breaking field is not scheduled.  Its coefficient is the
constant $\eta$ at every step, so all $s$-dependence still appears through
$f$ alone and boundary cancellation is unaffected.

\emph{Encoding each step.}
At each discrete step, $H_s$ is a linear combination of unitaries in the
LCU index set of the block encoding extended by the
degeneracy-breaking field: kinetic coefficients $c_u(x)$ for
$u\in T$ are unchanged, the $D$ term has total coefficient $\eta$
(since $D$ is unit-normalized and configuration-independent), and the potential
coefficients become $f(s)\,c_u(x)$ for $u\notin T\cup D$.
The normalization factor is therefore
$\lambda(x) = \lambda_T(x) + \eta
  + f(s)\sum_{u\notin T\cup D}\lvert c_u(x)\rvert$,
which at $f=1$ equals the full block-encoding normalization $\lambda(x)$ plus
$\eta$.
The block-encoding unitary is
$\mathsf U_{H_s}(x)
  = \mathsf{PREP}(x)^\dagger\,\mathsf{SEL}(x)\,\mathsf{PREP}(x)$,
where $\mathsf{SEL}$ is the $\mathsf{SEL}$ of the block encoding
extended by the $D$ term (momentum-diagonal phase unitaries of the
kinetic type, built by the same
$\mathsf{QFT}^\dagger\!\cdot\mathrm{phase}\cdot\mathsf{QFT}$
construction, fixed across configurations).  We modify $\mathsf{PREP}$ to
include the $D$ amplitudes weighted by $\eta$ (unit
normalization, $\textstyle\sum_{u\in D}\lvert c_u\rvert = 1$) and the
potential amplitudes scaled by $f(j/(m_d-1))$:
\begin{equation}\label{eq:prep-s}
\begin{split}
  &\mathsf{PREP}\,\lvert x\rangle\lvert 0\rangle_u
  = \lvert x\rangle
  \biggl(
    \sum_{u\in T}
      \sqrt{\frac{\lvert c_u(x)\rvert}{\lambda(x)}}\,\lvert u\rangle \\
  &\qquad
    + \sum_{u\in D}
      \sqrt{\frac{\eta\,\lvert c_u\rvert}{\lambda(x)}}\,
      \lvert u\rangle
    + \sum_{u\notin T\cup D}
      \sqrt{\frac{f(s)\,\lvert c_u(x)\rvert}{\lambda(x)}}\,
      \lvert u\rangle
  \biggr).
\end{split}
\end{equation}
Each step is the same circuit with different precomputed classical
constants.  There is no schedule register and no $\lambda$ register
($\lambda$ appears only inside the amplitude arithmetic).
As in the block encoding, all arithmetic, QROM lookups, and flag
controls are configuration-controlled, so $\mathsf U_{H_s}$ is block-diagonal
in $x$ and encodes $H_s(x)/\lambda(x)$ for each configuration.

\emph{Discrete walk operator.}
The walk operator at step $j$ is the matrix
exponential
\begin{equation}\label{eq:walk}
  \mathsf W_d(x) = \exp\!\bigl(-i\,H_s(x)/\lambda(x)\bigr),
\end{equation}
the first-order exponential integrator at time step $h=1$ applied to the
normalized operator.
The normalization gives $\lVert H_s/\lambda\rVert\le 1$, so the integrator's
validity condition $h \le 1/\lVert H_s/\lambda\rVert$ is satisfied at $h=1$.
This is the largest step admissible uniformly across configurations.  The
eigenphases of $\mathsf W_d$ occupy an arc of length at most
$2 < 2\pi$, so no eigenphase wrap-around occurs and the spectral gap of
$\mathsf W_d$ equals the gap of the normalized
Hamiltonian.

Normalization preserves the eigenpath.  Dividing by $\lambda(x)>0$ rescales
eigenvalues but leaves all eigenvectors unchanged on each configuration.
The followed eigenpath of $H_s/\lambda(x)$ is therefore that of the degeneracy-broken
Hamiltonian, and the endpoint eigenstate is the target eigenstate of
$\widetilde H(x) + \eta D$.

We construct the walk operator by applying a QSVT phase
sequence~\cite{gilyenQuantumSingularValue2019} to $\mathsf U_{H_s}$ with
precomputed phase angles, producing a block encoding of $\mathsf W_d$
that uses $\mathcal O(\log(m_d/\epsilon))$ queries to $\mathsf U_{H_s}$
at each step.
Because the encoded operator is already normalized, the phase angles are
universal constants independent of $x$ and $\lambda$.
We take an error allocation of order $\epsilon/m_d^2$ for each step (the conservative
accounting for a product of block encodings).  It keeps the total
walk-construction error within $\epsilon$, leaving the number of queries
$\mathcal O(\log(m_d/\epsilon))$ at each step.

\emph{Convergence.}
We apply the discrete adiabatic theorem with explicit gap
dependence~\cite{costaOptimalScalingQuantum2022}.
For a sequence of slowly varying gapped walk operators, the leakage of the
product $\prod_{j=0}^{m_d-1}\mathsf W_d(j/(m_d-1))$ out of the followed
eigenstate is bounded with explicit dependence on the walk operators'
finite-difference bounds and the multistep gap.
The discrete adiabatic theorem requires only that its hypotheses hold, not
an affine path.  The normalized interpolation $s\mapsto H_s/\lambda(x)$, with
its $s$-dependent normalization, is not affine.
We verify its hypotheses as follows.

(i)~\emph{Gap.}
Since $\lVert H_s/\lambda\rVert\le 1$, the walk gap equals the normalized
Hamiltonian gap with no eigenphase wrap-around.  Assumption~\ref{as:gap}
therefore transfers directly to $\mathsf W_d$.
(ii)~\emph{Finite differences.}
All $s$-dependence appears through $f$, since $\eta D$ is constant in $s$, and the $f$-derivative of
$(T + \eta D + f V)/(\lambda_T + \eta
  + f\sum_{u\notin T\cup D}\lvert c_u\rvert)$
is bounded in norm by a constant depending only on the coefficient ratio of
assumption~\ref{as:domain}.
With $\sup\lvert f'\rvert$ and $\sup\lvert f''\rvert$ bounded, the first
and second finite differences of $\mathsf W_d$ are
$\mathcal O(1/m_d)$ and $\mathcal O(1/m_d^2)$, respectively, with explicit
constants determined by these bounds.
The theorem's boundary terms (finite-difference values at the
endpoints, nonzero for any non-constant schedule) do not worsen the
rigorous scaling and are strongly suppressed by the glue schedule's
endpoint flatness.
(iii)~\emph{Boundary cancellation (derivative-level).}
By the chain rule, every $s$-derivative of $\mathsf W_d$ contains a
derivative of $f$ as a factor, so the glue schedule's vanishing endpoint
derivatives ($f^{(k)}(0) = f^{(k)}(1) = 0$ for all $k\ge 1$) transfer to
the walk.
This derivative-level cancellation is distinct from the finite-difference
boundary terms of~(ii).  It is the condition that the high-order tier
uses.

Block-diagonality in $x$ means the $m_d$-step evolution acts independently
on each configuration.
The theorem applies configuration by configuration.  The uniform $\Delta$ covers the worst-case configuration, and the leakage bound
holds for every configuration simultaneously.  The superposition satisfies the
target-eigenstate guarantee to within $\epsilon$ (adiabatic leakage plus
walk-construction error).

We give three tiers of scaling.

\emph{Tier~1 (rigorous).}
The same discrete adiabatic
theorem, with the finite-difference bounds of~(ii) and the uniform gap, gives
$m_d = \mathcal O((\lambda/\Delta)^{3}\epsilon^{-1})$
at $h=1$, with constants that are uniform
and depend only on the schedule derivatives and the coefficient ratio.
This is the same scaling as the analogous affine-path result.

\emph{Tier~2 (conditional).}
Under assumption~\ref{as:highorder}, the high-order discrete adiabatic
theorem applied to our walk (boundary cancellation supplied by~(iii),
constant-level gap, $h=1$) gives
$m_d = \mathcal O(\epsilon^{-1/k})$ for every $k\ge 1$, i.e.\
$\mathcal O(1/\epsilon^{o(1)})$, the same form as the analogous affine-path
result.

\emph{Tier~3 (anticipated practical scaling).}
In the continuous setting, boundary-cancellation schedules of Gevrey class
$\alpha$ give evolution time
$\mathcal O((\lambda/\Delta)^{2}\log^{1+\alpha}(\lambda/(\Delta\epsilon)))$.
For the glue schedule ($\alpha=1$ by a standard Gevrey estimate), the
anticipated discrete scaling is
$m_d = \mathcal O((\lambda/\Delta)^{2}\log^{2}(\lambda/(\Delta\epsilon)))$, consistent with the numerical
evidence for exponential convergence.
A discrete-time proof would follow from the high-order discrete adiabatic
theorem, currently a conjecture.  We use this tier in the resource estimates,
since the numerical evidence supports its scaling while Tier~1 reflects
the present proof.

We build every step of the evolution from the flag-controlled encoding, so the walk preserves the image of the active-electron embedding
and the isolation acts entirely within it.

\emph{Remarks.}
Each configuration's electronic state is the $\eta$-selected target eigenstate
of $\widetilde H(x) + \eta D$ up to a relative phase, to within $\epsilon$
(adiabatic leakage plus walk-construction error), with the block-encoding
ancillas restored to $\lvert 0\rangle$ within the same error allocation.
If a residual accidental degeneracy survives the breaking for some configuration,
assumption~\ref{as:gap} fails there and the simple-eigenstate guarantee is
unavailable.
Exact symmetry-protected degeneracies whose partners the path never couples
(Kramers doublets at odd $N_e$: the interpolation varies only through the
time-even $f(s)V$, and time-even terms have vanishing matrix elements
between partners) leave the
followed branch effectively isolated, with the operative gap set by the
non-target spectrum.

The property primitives evaluate the implemented $\widetilde H(x)$.  Its eigenvalues match those of
$H(x)$ to within $\lVert\widetilde H - H\rVert$.
The prepared state lies within
$\mathcal O(\eta) + \mathcal O(\lVert\widetilde H - H\rVert/\Delta)$ of
the target eigenspace of $H(x)$ in norm, so the
phase-estimation outcome is supported on the target eigenvalue up to an
$\mathcal O(\eta^2 + \lVert\widetilde H - H\rVert^2/\Delta^2)$ probability
on other eigenvalues, an error contribution we allocate alongside the
isolation leakage.
Expectation-valued properties have the first-order $\mathcal O(\eta)$
selection bias, following the $\eta$ trade-off stated above.

We deliberately omit each configuration's phases from the guarantee.  Because the
isolation is one fixed unitary, applied and later uncomputed as a whole,
these phases cancel identically and need not be tracked.
Our construction, started from the corresponding excited eigenstate of
$T + \eta D$ (also a degeneracy-broken plane-wave Slater determinant), isolates
the corresponding eigenstate of $\widetilde H(x) + \eta D$ under
assumption~\ref{as:gap} for that eigenpath.
The isolated state is the input to the spectral primitive.

\section{Resource Estimation}\label{app:resource-estimation}

We estimate the number of logical qubits and the logical Toffoli depth of
the full search circuit, first as asymptotic bounds in terms of a
single tolerance~$\epsilon$, then as explicit integer-rounded formulas for the case studies.  Costs cover the quantum computation only.
We exclude classical relaxation and verification after measurement.

\emph{Assumptions.}
We apply the following conventions throughout.
(\refstepcounter{asmpRE}\theasmpRE\label{as:re-tol})~\emph{Tolerances.}
The algorithmic tolerance $\epsilon$ covers the arithmetic, primitive,
derived-property, and
constraint tolerances.  It is the spectral primitive's error allocation,
separating into systematic shifts (the encoding error of
\cref{eq:spectral-transfer}, the $\mathcal O(\eta)$ selection bias, the phase
resolution, and finite-bit arithmetic) and leakage (the isolation leakage, the
$\mathcal O(\eta^2)$ off-target probability, and the taper failure probability,
whose average-case guarantee~\cite{patelOptimalCoherentQuantum2026} transfers to the configuration mixture by
linearity).  We compose this allocation across properties and collapse it to its
dominant component.
Constraint thresholds are fixed case-study constants, with the pass
tolerance we choose below them.  We take real-arithmetic registers to have
$b=\mathcal O(\log(1/\epsilon))$ bits.  We fix the number of sampled-potential phases
by the sampling-error bound (it scales
with the sampled-potential 1-norm, hence with system size and the potential
maxima, divided by $\epsilon$) and it appears only through the LCU index and selection
registers, where it is sub-dominant to the arithmetic.  It is an intermediate
quantity and is absent from the system-variable formulas below.
(\refstepcounter{asmpRE}\theasmpRE\label{as:re-geom})~\emph{Bounded cell geometry.}
The allowed unit cell structure keeps $\mathbf G$ well-conditioned, with bounded
derivatives, and $[\det\mathbf G]^{-1/6}$ bounded by a constant
independent of $N$, $N_e$, and $\epsilon$.
(\refstepcounter{asmpRE}\theasmpRE\label{as:re-chan})~\emph{Bounded pseudopotential channels.}
Channels, angular momenta, the number of projectors, and table-entry magnitudes are
bounded constants, and the coefficients are smooth in the encoded
structural parameters with bounded
derivatives.
(\refstepcounter{asmpRE}\theasmpRE\label{as:re-parallel})~\emph{Parallel logical layer.}
We take depth as logical Toffoli depth under the standard logical-layer abstraction:
unrestricted parallel Toffoli execution, all-to-all connectivity, Clifford
operations (including fanout) not counted toward depth.  We take the Toffoli counts cited for each call
as conservative depth bounds, since depth never exceeds
count.  We count explicitly parallel structures (pair arithmetic, tree
summation) separately.
(\refstepcounter{asmpRE}\theasmpRE\label{as:re-qrom})~\emph{QROM convention.}
Species lookup over $M$ entries uses the select-swap QROM
construction (Theorem~2 of~\cite{berryQubitizationArbitraryBasis2019}).  Under the
parallel-logical-layer convention above, our analysis uses the
parallelized setting, lookup depth logarithmic in $M$ at the cost of $\mathcal O(M)$
ancillas per active primitive.
(\refstepcounter{asmpRE}\theasmpRE\label{as:re-prep})~\emph{Uniform preparation.}
Uniform $c_x$ over the allowed configuration domain and a simple electronic initial state form the baseline of our estimate.
The latter is the single fixed initial determinant,
valid because the allowed cell domain holds the occupied plane-wave set
fixed across configurations.  Arbitrary-amplitude preparation routines are out of
scope.  The initial-state reflection is consequently lower order than the flag
function.
(\refstepcounter{asmpRE}\theasmpRE\label{as:re-branch})~\emph{Eigenstate preparation for each configuration.}
The number of primitives evaluated per configuration, $N_P$, divides into the numbers of spectral, static,
and dynamical primitives, $N_P = N_{P,\mathrm{spec}} + N_{P,\mathrm{stat}} + N_{P,\mathrm{dyn}}$.
Each active primitive has its own electronic register and an isolated target eigenstate, prepared by state-isolation stages
acting on all $N_P$ registers in parallel.  The ancilla registers are
$N_P$-fold (counted in the qubit total), but we count the isolation depth once.
$N_P$ therefore multiplies active ancilla registers but not
primitive-evaluation depth.
(\refstepcounter{asmpRE}\theasmpRE\label{as:re-pqgap})~\emph{Projected queries.}
Static and dynamical estimation act on the target eigenstate prepared by the
isolation stage.  Each query applies an eigenstate filter, a polynomial of degree
$\mathcal O(\lambda/\Delta)$ in the block encoding, that returns the
state to the target eigenspace.  We suppress the logarithmic
tolerance/success factors in the asymptotic layer and
absorb them into the explicit channel-error factor in the case-study formulas.
(\refstepcounter{asmpRE}\theasmpRE\label{as:re-const})~\emph{Bounded construction constants.}
The degeneracy-breaking field, the minimum-image search over the $3^3$ images, and the
adjoint-index swap are bounded or constant-factor contributions.  They do not affect the
asymptotic exponents.

\subsection{Qubit Count}\label{sec:re-qubits}

\emph{Configuration state.}
The species, unit-cell, and fractional-coordinate registers
($\mathbf Z$, $\mathbf a$, $\mathbf R$) at $b$ bits per real contribute
\begin{equation}\label{eq:q-config}
  \mathcal O\!\left(N\log M + N\log\!\left(\frac{1}{\epsilon}\right)\right).
\end{equation}

\emph{Primitive ancilla registers for each configuration.}
One active primitive stores electronic coordinates and spins, activity flags,
pair-distance and single-particle work registers, LCU/selection ancillas, and QROM
ancillas:
\begin{equation}\label{eq:q-branch}
\begin{split}
  \mathcal O\biggl(&M + (N_e{+}N)^2\log\!\left(\frac{1}{\epsilon}\right) \\
    &+ \log M + \log(N_e{+}N)
    + \log\!\log\!\left(\frac{1}{\epsilon}\right)\biggr),
\end{split}
\end{equation}
multiplied by $N_P$ under the preparation convention for each configuration.  The
$(N_e{+}N)^2\log(1/\epsilon)$ term counts the pairwise distance/potential
ancillas.  The $M$ term counts the select-swap QROM ancillas.

\emph{Primitive-value and derived-property registers.}
These contribute $\mathcal O((N_P + N_{\mathrm{prop}})\log(1/\epsilon))$, with
$N_{\mathrm{prop}}$ the number of target properties.  These registers are counted
separately from the primitive ancilla registers.

\emph{Total.}
We combine the three contributions, producing the number of qubits
\begin{equation}\label{eq:q-total}
\begin{split}
  \mathcal O\biggl(&N_P M + (N{+}N_P)\log M \\
  &+ \bigl[N + N_P + N_{\mathrm{prop}}
    + N_P(N_e{+}N)^2\bigr]
    \log\!\left(\frac{1}{\epsilon}\right)\biggr).
\end{split}
\end{equation}

\subsection{Toffoli Depth}\label{sec:re-depth}

\emph{Normalization.}
$\lambda$ follows from the LCU coefficients and bounded cell geometry, with the grid cutoff
fixed by the tolerance.  Kinetic coefficients
scale as momentum squared, $\mathcal O(N_e/\epsilon^2)$.  Truncated
electron--electron and nucleus--nucleus Coulomb terms scale with the
tolerance, $\mathcal O((N_e{+}N)^2/\epsilon)$.  Local and nonlocal
pseudopotential terms are bounded-channel sums, $\mathcal O(N_e N)$.  Together,
\begin{equation}\label{eq:lambda-scale}
  \lambda = \mathcal O\!\left(
    \frac{N_e}{\epsilon^2}
    + \frac{(N_e{+}N)^2}{\epsilon}
    + N_e N\right).
\end{equation}
We use the uniform bound
$\lambda_{\max} = \max_x \lambda(x)$ taken over the searched class.  The search
circuit has one fixed size, set by the worst-case configuration.  Bounded cell geometry and bounded channels keep this bound at the same
order as the typical configuration's value, and subsequent formulas write $\lambda$ for this
bound.

\emph{Block encoding.}
One $\mathsf U_H(x)$ call consists of metric arithmetic, species QROM lookup,
normalization arithmetic, $\mathsf{PREP}$, and $\mathsf{SEL}$ arithmetic, in the first-quantized arithmetic
style of~\cite{suFaultTolerantQuantumSimulations2021}.  The cost of each call is
$\mathcal O(b^2)$ Toffolis per fixed-point operation.  With pair arithmetic in
parallel, tree summation, and the QROM convention, the depth of a single call is
\begin{equation}\label{eq:DH}
  D_H = \mathcal O\!\left(\log M
    + \log\!\left(\frac{1}{\epsilon}\right)
      \log\!\left(\frac{N_e{+}N}{\epsilon}\right)\right).
\end{equation}
Sampled-potential selection adds $\mathcal O(\log(1/\epsilon))$ space and
$\mathcal O(\log^2(1/\epsilon))$ depth, which we absorb into $D_H$.

\emph{State isolation.}
We divide the leakage allocation across the $\mathcal O(N_P/\sqrt{\rho})$
isolation uses of the search, so each use has the target
$\mathcal O(\sqrt{\rho}\,(1-p)\,\epsilon/N_P)$.
From our isolation analysis: $m_d = \mathcal O((\lambda/\Delta)^{2}\log^2(N_P\lambda/(\sqrt{\rho}(1-p)\Delta\epsilon)))$ discrete steps
under the Tier-3 scaling reported there, each step requiring
$\mathcal O(\log(N_P\lambda/(\sqrt{\rho}(1-p)\Delta\epsilon)))$ QSVT-block calls, applied to
all $N_P$ primitives in
parallel:
\begin{equation}\label{eq:Diso}
\begin{split}
  D_{\mathrm{iso}}
  &= \mathcal O\!\left(m_d\log\!\left(\frac{N_P\,\lambda}{\sqrt{\rho}\,(1-p)\,\Delta\epsilon}\right)D_H\right) \\
  &= \mathcal O\!\left(\left(\frac{\lambda}{\Delta}\right)^{2}\log^3\!\left(\frac{N_P\,\lambda}{\sqrt{\rho}\,(1-p)\,\Delta\epsilon}\right)D_H\right).
\end{split}
\end{equation}
The degeneracy-breaking term $\eta D$ is one unit-normalized one-body LCU term, which we absorb
in $D_H$.

\emph{Spectral.}
Tapered QPE on the walk operator~\cite{patelOptimalCoherentQuantum2026}
requires $m_q = \mathcal O(\lambda/\epsilon)$ walk queries at absolute
tolerance~$\epsilon$ (rms convention).  We absorb the logarithmic taper
failure-suppression factor
(Theorem~3 of~\cite{patelOptimalCoherentQuantum2026}).  Substituting $\lambda$:
\begin{equation}\label{eq:Dspec}
\begin{split}
  D_{\mathrm{spec}}
  &= \mathcal O\!\left(\frac{\lambda}{\epsilon}\,D_H\right) \\
  &= \mathcal O\!\left(\frac{1}{\epsilon}\!\left[
    \frac{N_e}{\epsilon^2}
    + \frac{(N_e{+}N)^2}{\epsilon}
    + N_e N\right]D_H\right),
\end{split}
\end{equation}
up to lower-order phase and energy arithmetic.

\emph{Static.}
We estimate the resources of the static and dynamical primitives here.
Bounded dimensionless operators (e.g.\ local spin) have $\mathcal O(1)$ estimation
scale.  A static primitive estimates the Hellmann--Feynman derivative
$\langle\psi_n|\partial H/\partial q|\psi_n\rangle$ for a generalized
parameter~$q$, an encoded fractional coordinate, unit-cell edge length, or
angle.  The Hamiltonian $H = \sum_u c_u\,\mathsf U_u$ depends on~$q$ through
both the coefficients $c_u$ and the data-controlled unitaries $\mathsf U_u$.
Both are piecewise smooth in~$q$, with sharpest
variation scale equal to the tolerance $\epsilon$, set
by the truncation and minimum-image constructions.
We then bound the derivative's block-encoding
normalization by the product rule:
\begin{equation}\label{eq:deriv-scale}
  \sum_u \left|\frac{\partial c_u}{\partial q}\right|
  + \sum_u \lvert c_u\rvert
    \left\|\frac{\partial \mathsf U_u}{\partial q}\right\|
  = \mathcal O\!\left(\frac{\lambda}{\epsilon}\right),
\end{equation}
giving a static estimation scale of $\mathcal O(\max\{1,\lambda/\epsilon\})$.  This is
our conservative bound tied to assumptions~\ref{as:re-geom}--\ref{as:re-chan} and the
encoded-coordinate normalization.  It requires the configuration to lie in a
smooth region, away from the nonsmooth points of the construction (minimum-image
switches, reduced-cell boundaries, the Coulomb truncations, and the $V_+/V_-$ sign
changes).  At those points we take the derivative as a one-sided subgradient
finite difference at the same scale.  Derivatives with respect to unnormalized
physical coordinates, and pseudopotentials with unbounded derivative constants,
lie outside the scope of this bound.

Coherent iterative amplitude
estimation is non-destructive and
SVT-based, with a number of queries that scales, up to a logarithmic channel-error
factor, inversely with the normalized target tolerance
(Corollary~20 of~\cite{rallFasterCoherentQuantum2021}).
At normalized tolerance
$\epsilon\cdot(\epsilon/\lambda)$, with each query applying the eigenstate
filter, the estimation gives
\begin{equation}\label{eq:Dstat}
  D_{\mathrm{stat}}
  = \mathcal O\!\left(\frac{\lambda}{\Delta\epsilon^2}\!\left[
    \frac{N_e}{\epsilon^2}
    + \frac{(N_e{+}N)^2}{\epsilon}
    + N_e N\right]D_H\right).
\end{equation}
For bounded dimensionless static primitives the factor is
$\mathcal O(\lambda/(\epsilon\Delta))$ instead.  We count the global isolation
stage once, in $D_{\mathrm{iso}}$, and do not repeat it inside the estimation loop.

\emph{Dynamical.}
The broadening $\eta$ is a physical energy in the same units as~$\epsilon$.
It sets the resolution of the computed response.  The estimated quantity is
the $\eta$-scaled response (the resolvent multiplied by~$\eta$, a bounded
operator), so the absolute tolerance~$\epsilon$ applies directly.  On the
encoded $\widetilde H/\lambda$ spectral scale the resolvent width is
$\eta/\lambda$, with $0<\eta\le\lambda$.

The resolvent circuit requires the reference eigenphase (encoding the
reference energy that shifts the resolvent) to be available in its own
storage register before the parallel primitive layer completes.  A
reference error displaces each of the $r$ resolvent factors and shifts the
$\eta$-scaled response by the ratio of that error to $\eta$, so the reference must be
known to $\mathcal O(\epsilon\eta/r)$.  The spectral estimate at this
tolerance uses $\mathcal O(r\lambda/(\epsilon\eta))$ queries, counted in
\eqref{eq:Ddyn} as the $D_{\mathrm{spec}}$ term with the remainder absorbed
by the resolvent term.

We implement the resolvent by quantum signal processing or quantum singular value transformation frameworks.  Our estimate uses generalized quantum signal processing (GQSP), an established algorithm~\cite{motlaghGeneralizedQuantumSignal2024}.  An $\epsilon$-accurate polynomial of the rescaled inverse on a domain bounded
away from zero by $\eta/\lambda$ has degree
$\mathcal O((\lambda/\eta)\log(1/\epsilon))$~\cite{gilyenQuantumSingularValue2019}.
GQSP implements a polynomial of the walk with a number of controlled queries
equal to the polynomial degree.  For $r$ resolvent factors, where $r$ is the
response order, we split the tolerance across the factors, giving the total
degree $\mathcal O((r\lambda/\eta)\log(r/\epsilon))$.

We repeat this resolvent block and
the eigenstate filter $\mathcal O(1/\epsilon)$ times by coherent iterative amplitude estimation:
\begin{equation}\label{eq:Ddyn}
  D_{\mathrm{dyn}}
  = \mathcal O\!\bigl(D_{\mathrm{spec}}\bigr)
  + \mathcal O\!\left(\frac{1}{\epsilon}\!\left[\frac{r\lambda}{\eta}
    \log\!\left(\frac{r}{\epsilon}\right)
    + \frac{\lambda}{\Delta}\right]D_H\right).
\end{equation}
The filter term is sub-dominant to the resolvent term when
$\eta\le\Delta$, up to logarithmic factors.
Choices of $f(\hat H)$ that are bounded and smoother than the resolvent may cost less than this
bound.  We estimate perturbation/operator blocks conservatively at $\mathcal
O(D_H)$.

\emph{Search.}
We take the parameter $\rho$ as a lower bound on the solution density seen by
the prepared search state (uniform enumerated case $\rho = N_{sol}/N_x$, with
$N_{sol}$ the number of solution configurations and $N_x$ the number of
configurations in the search space).
The fixed-point construction tolerates underestimates, since the success guarantee
only improves when the actual density exceeds the
bound (Eq.~(2) of~\cite{yoderFixedPointQuantumSearch2014}).  Fixed-point search reaches
success probability~$p$ in
\begin{equation}\label{eq:search-m}
  m = \left\lceil\frac{1}{2}\left(
    \frac{\ln\!\bigl(\tfrac{2}{\sqrt{1-p}}\bigr)}{\sqrt{\rho}}
    - 1\right)\right\rceil
\end{equation}
iterations (Eq.~(3) of~\cite{yoderFixedPointQuantumSearch2014}).
Each iteration comprises flag-function evaluation, the phase reflection, flag-function uncomputation, and
the initial-state reflection (lower order under uniform preparation).  The
primitive-to-property arithmetic, comparison, and Boolean layer adds
$\mathcal O((N_P + N_{\mathrm{prop}})\log(1/\epsilon))$ depth per flag-function call
(linear-size reversible arithmetic, not parallelized in this layer).

\emph{Total.}
Write $[\,\cdot\,]$ for the Iverson bracket, equal to~$1$ when the enclosed
condition holds and~$0$ otherwise.  The assembled depth bound is
\begin{equation}\label{eq:depth-total}
\begin{aligned}
  \mathcal O\Biggl(
    \frac{\ln\!\bigl(\tfrac{2}{\sqrt{1-p}}\bigr)}{\sqrt{\rho}}
    \Bigl(
      &D_{\mathrm{iso}}
      + (N_P + N_{\mathrm{prop}})\log\!\left(\tfrac{1}{\epsilon}\right) \\
      &+ \max\bigl\{
        [N_{P,\mathrm{spec}}{>}0]\,D_{\mathrm{spec}}, \\
      &\qquad\quad
        [N_{P,\mathrm{stat}}{>}0]\,D_{\mathrm{stat}}, \\
      &\qquad\quad
        [N_{P,\mathrm{dyn}}{>}0]\,D_{\mathrm{dyn}}
      \bigr\}
    \Bigr)
  \Biggr).
\end{aligned}
\end{equation}
Under primitive parallelism the depth is the maximum over active classes,
while only the ancilla registers depend on the number of primitives in each class.

\subsection{Case Study Formulas}\label{sec:re-case}

We convert the asymptotic bounds above into deterministic, integer-rounded
formulas $Q_{\mathrm{case}}$ and $D_{\mathrm{case}}$
for the case studies.  We fill in each $\mathcal O(\cdot)$ block with
explicit numbers of registers and representative arithmetic allowances.  Our estimates
are order-of-magnitude.

\emph{Arithmetic allowances.}
The Toffoli count for each call bounds the depth.  Component allowances: fixed-point
multiplication $\sim\!\tfrac{3}{2}b^2$ and inverse square root
$\sim\!\tfrac{15}{2}b^2$ per Newton iteration, a few iterations being
enough~\cite{hanerOptimizingQuantumCircuits2018}; $\mathcal O(b^2)$ per
first-quantized arithmetic
operation; and the select-swap
QROM lookup at logarithmic depth.
We assign the block-encoding call the conservative direct-evaluation
estimate $105\,b^2$: one reciprocal/root allowance ($30b^2$), six bounded
smooth-function allowances ($72b^2$), one multiplication ($2b^2$), and one tree
reduction ($b^2$).  We take this as a deliberate upper bound on the unoptimized direct-evaluation path.  The
first-quantized construction often avoids direct evaluation through algebraic
rearrangement, and a different
arithmetic library changes the constant, not the structure.

\emph{Tolerance inputs.}
The pass tolerance~$\epsilon$ (the coarse pass uses
$\epsilon_{\mathrm{coarse}}$, the fine pass $\epsilon_E$) is the primary
input.  We introduce the grid tolerance $\epsilon_{\mathrm{grid}}$ as an
explicit input.  It is a discretization-validity parameter.
Because the primitive estimates the discretized Hamiltonian's spectrum, the
grid error must lie below the requested tolerance.  Under the smooth
wavefunction assumption of the HGH
pseudopotential,
we relax the conservative
$\epsilon_{\mathrm{grid}} = \epsilon$ bound to the rule by second-order grid convergence
\begin{equation}\label{eq:grid-rule}
  \epsilon_{\mathrm{grid}} \le \sqrt{\epsilon}\,,
\end{equation}
with $\epsilon$ the pass tolerance.  We choose a grid for each pass.

\emph{Qubit count.}
The explicit integer-rounded formula is
\begin{equation}\label{eq:Qcase}
\begin{aligned}
  Q_{\mathrm{case}}(\epsilon)
  ={}& N\lceil\log_2(\max\{2,M\})\rceil
    + (6+3N)\lceil\log_2(1/\epsilon)\rceil \\
  &+ N_P\Bigl[
    3N_e\lceil\log_2(1/\epsilon)\rceil + 2N_e \\
  &\quad
    + 6\lceil\log_2(1/\epsilon)\rceil
      \bigl(\tfrac{N_e(N_e{-}1)}{2} \\
  &\qquad\qquad
    + N_e N + \tfrac{N(N{-}1)}{2}\bigr) \\
  &\quad
    + 3(N_e{+}N)\lceil\log_2(1/\epsilon)\rceil \\
  &\quad
    + \lceil\log_2(\max\{2,M\})\rceil \\
  &\quad
    + M + 2\lceil\log_2(\max\{2,N_e{+}N\})\rceil \\
  &\quad
    + 4\lceil\log_2(1/\epsilon)\rceil \\
  &\quad
    + \bigl\lceil\log_2\bigl(
        \max\{2,\lceil\log_2(1/\epsilon)\rceil\}\bigr)\bigr\rceil
    + 4\Bigr] \\
  &+ (1 + N_P + N_{\mathrm{prop}})\lceil\log_2(1/\epsilon)\rceil \\
  &+ (N_{\mathrm{prop}} + 1).
\end{aligned}
\end{equation}
The lines follow the qubit-count subsection: the configuration state;
the electronic, pair, single-particle, lookup, and selection ancilla registers for each configuration (multiplied
by $N_P$); and the value and property registers.

\emph{Toffoli depth.}
We evaluate the depth symbols as explicit closed forms, reusing the number of search iterations
\begin{equation}\label{eq:mcase}
  m = \left\lceil\frac{1}{2}\left(
    \frac{\ln\!\bigl(\tfrac{2}{\sqrt{1-p}}\bigr)}{\sqrt{\rho}}
    - 1\right)\right\rceil,
\end{equation}
with $\rho$ and $p$ the pass's solution-density lower bound and success
probability, together with the integer form of \eqref{eq:lambda-scale}, taken
with exact numbers of pairs,
\begin{equation}\label{eq:lambda-case}
  \lambda = \frac{N_e}{\epsilon_{\mathrm{grid}}^2}
    + \frac{\tfrac{N_e(N_e{-}1)}{2} + \tfrac{N(N{-}1)}{2}}
           {\epsilon_{\mathrm{grid}}}
    + N_e N.
\end{equation}
The block-encoding depth uses the conservative $105\,b^2$ estimate above,
and the primitive-class depths reuse $\lambda$, $D_H$, and $m$:
\begin{equation}\label{eq:DHcase}
\begin{split}
  D_H ={}& 2\lceil\log_2(\max\{2,M\})\rceil
    + 105\lceil\log_2(1/\epsilon)\rceil^2 \\
  &+ \lceil\log_2(1/\epsilon)\rceil
      \lceil\log_2(\max\{2,N_e{+}N\})\rceil \\
  &+ \lceil\log_2(\max\{2,N_e{+}N\})\rceil
    + \lceil\log_2(1/\epsilon)\rceil.
\end{split}
\end{equation}
\begin{equation}\label{eq:Disocase}
\begin{split}
  D_{\mathrm{iso}} ={}& \left\lceil
    \left(\frac{\lambda}{\Delta}\right)^{2}
    \left\lceil\log_2\!\left(
      \max\left\{2,\frac{\lambda}{\Delta\epsilon}\,
        \frac{\max\{2,2N_P m\}}{1-p}\right\}
    \right)\right\rceil^{2}\right\rceil \\
  &\times \left\lceil\log_2\!\left(
      \max\left\{2,\frac{\lambda}{\Delta\epsilon}\,
        \frac{\max\{2,2N_P m\}}{1-p}\right\}
    \right)\right\rceil D_H.
\end{split}
\end{equation}
\begin{equation}\label{eq:Dspeccase}
  D_{\mathrm{spec}}
  = \left\lceil\frac{\lambda}{\epsilon}\right\rceil D_H
    + 36\lceil\log_2(1/\epsilon)\rceil^2.
\end{equation}
\begin{equation}\label{eq:Dstatcase}
\begin{split}
  D_{\mathrm{stat}} ={}&
    \left\lceil\frac{\lambda}{\epsilon\,\epsilon_{\mathrm{grid}}}\right\rceil
    \left\lceil\frac{\lambda}{\Delta}\right\rceil \\
  &\times
    \left\lceil\log_2\!\left(\frac{\max\{2,2N_P m\}}{1-p}\right)\right\rceil
    D_H \\
  &+ 3\lceil\log_2(1/\epsilon)\rceil.
\end{split}
\end{equation}
\begin{equation}\label{eq:Ddyncase}
\begin{split}
  D_{\mathrm{dyn}} ={}& D_{\mathrm{spec}} \\
  &+ \biggl\lceil\frac{1}{\epsilon}\biggl(\frac{r\lambda}{\eta}
      \left\lceil\log_2\!\left(\frac{r}{\epsilon}\right)\right\rceil
      + \left\lceil\frac{\lambda}{\Delta}\right\rceil\biggr)
      \biggr\rceil \\
  &\times
    \left\lceil\log_2\!\left(\frac{\max\{2,2N_P m\}}{1-p}\right)\right\rceil
    D_H \\
  &+ 36\lceil\log_2(1/\epsilon)\rceil^2.
\end{split}
\end{equation}
The $36\lceil\log_2(1/\epsilon)\rceil^2$ terms are phase/energy arithmetic, the
$3\lceil\log_2(1/\epsilon)\rceil$ term is value arithmetic, and the
$\lceil\log_2(\max\{2,2N_P m\}/(1-p))\rceil$ factor is the coherent-estimation
channel-error/success-amplification allocation across all primitive uses in the
search.  The isolation logarithms include the same allocation.  The broadening $\eta$ is in the same units as
the pass tolerance.  With $N_{P,\mathrm{dyn}} = 0$, $\eta$ and $r$ are not
inputs.

The composed depth is
\begin{equation}\label{eq:Dcase}
\begin{aligned}
  D_{\mathrm{case}}&(\epsilon,\rho,p,\Delta,\eta)
  = m\Bigl[ 2\Bigl( D_{\mathrm{iso}} \\
  &+ \max\bigl\{
      [N_{P,\mathrm{spec}}{>}0]\,D_{\mathrm{spec}}, \\
  &\qquad\quad
      [N_{P,\mathrm{stat}}{>}0]\,D_{\mathrm{stat}}, \\
  &\qquad\quad
      [N_{P,\mathrm{dyn}}{>}0]\,D_{\mathrm{dyn}}
      \bigr\} \\
  &+ 2(N_P + N_{\mathrm{prop}} + 1)\lceil\log_2(1/\epsilon)\rceil \\
  &+ \lceil\log_2(1/\epsilon)\rceil
      \lceil\log_2(\max\{2,N_{\mathrm{prop}}{+}1\})\rceil
    \Bigr) \\
  &+ \lceil\log_2(\max\{2,M\})\rceil
    + \lceil\log_2(\max\{2,N_e{+}N\})\rceil \\
  &+ \lceil\log_2(1/\epsilon)\rceil + 1\Bigr].
\end{aligned}
\end{equation}
We count the factor of two as the compute and uncompute halves of each iteration's work.
The trailing terms are the initial-state reflection.

\emph{Two-pass combination.}
We take a coarse search pass
$(\epsilon_{\mathrm{coarse}},\rho_{\mathrm{coarse}},p_{\mathrm{coarse}})$
to narrow the candidate set.  We take a fine pass
$(\epsilon_E,\rho_{\mathrm{fine}},p_{\mathrm{fine}})$ to verify the result ($\rho_{\mathrm{fine}} = 1$ for one
output configuration that meets the targets at the fine tolerance).  The two-pass
totals are
\begin{equation}\label{eq:twopass}
\begin{split}
  Q_{\mathrm{case}}^{(2)}
    &= \max\!\bigl\{Q_{\mathrm{case}}(\epsilon_{\mathrm{coarse}}),\,
       Q_{\mathrm{case}}(\epsilon_E)\bigr\}, \\
  D_{\mathrm{case}}^{(2)}
    &= D_{\mathrm{case}}(\epsilon_{\mathrm{coarse}},
       \rho_{\mathrm{coarse}},p_{\mathrm{coarse}},\Delta,\eta_{\mathrm{coarse}}) \\
    &\quad + D_{\mathrm{case}}(\epsilon_E,
       \rho_{\mathrm{fine}},p_{\mathrm{fine}},\Delta,\eta_{\mathrm{fine}}).
\end{split}
\end{equation}
\Cref{eq:twopass} combines the two single-pass estimates.  The fine pass contributes its full cost.

\emph{Solution density.}
Under exhaustive grid enumeration the configuration space contains at most
$N_{\mathbf Z} \le M^N$ species assignments,
$N_{\mathbf a} \lesssim \epsilon^{-6}$ cell grid points, and
$N_{\mathbf R} \lesssim \epsilon^{-3N}$ coordinate grid points, so
$N_x \lesssim M^N\epsilon^{-(3N+6)}$.
Suppose the constraint thresholds (including equality windows) are fixed case-study constants, we choose the pass tolerance below
them, and each solution configuration occupies a coordinate region whose volume fraction is bounded below (the bound we absorb into $N_{sol}$).  Then refining the coordinate grid scales $N_{sol}$ and $N_x$ identically.  The
grid factor cancels, and $\rho$ is grid-resolution-independent. $N_e$
never appears in the configuration space.  Throughout, $N_{sol}$ denotes
an effective, region-weighted lower-bound number of solutions,
consistent with $\rho$ being a lower bound in the search.
We tighten the species space by replacing $M^N$ with
$M_{\mathrm{eff}}^N$, where
$M_{\mathrm{eff}}\in[1,M]$ is, at each site, the effective number of species that pass
explicit local rules (charge neutrality, geometric tolerance,
valence/coordination compatibility, radius-ratio bounds), giving
\begin{equation}\label{eq:rho-decomp}
  \rho = \min\!\left\{1,\;
    N_{sol}\,M_{\mathrm{eff}}^{-N}\right\}.
\end{equation}
Treating sites independently over-counts solutions, but strict and loose rule sets bound the true cost from both sides (the search factor scales as $M_{\mathrm{eff}}^{N/2}$).  For emergent constraints not reducible to individual sites, $M_{\mathrm{eff}} \to M$.  This $\rho^{-1/2}$ factor sets only the number of search iterations, not the cost of each iteration.

\emph{Illustrative scaling points.}
\Cref{tab:scaling-points} lists the inputs at which we evaluate
\eqref{eq:Dcase} for the Toffoli-depth scaling in the main text.

\begin{table}[t]
\caption{Inputs for the three illustrative scaling points.  Shared:
  $p=0.95$, $\Delta=10^{-2}$~Ha, $\eta=5\times10^{-4}$~Ha, $r=1$, $M_{\mathrm{eff}}=1.5$,
  $\epsilon_{\mathrm{grid}}=\sqrt{\epsilon}$, $\rho=M_{\mathrm{eff}}^{-N}$.
  $\epsilon$ in Ha.}%
  \label{tab:scaling-points}
\centering
\begin{minipage}{0.7\columnwidth}
\begin{ruledtabular}
\begin{tabular}{lccc}
 & Small & Medium & Large \\
\colrule
$M$ & 2 & 5 & 10 \\
$N$ & 2 & 14 & 64 \\
$N_e$ & 4 & 100 & 700 \\
$N_{P,\mathrm{spec}}$ & 1 & 2 & 3 \\
$N_{P,\mathrm{stat}}$ & 1 & 12 & 80 \\
$N_{P,\mathrm{dyn}}$ & 1 & 12 & 40 \\
$N_{\mathrm{prop}}$ & 1 & 5 & 8 \\
$\epsilon$ & $10^{-2}$ & $10^{-2}$ & $3\times10^{-3}$ \\
\end{tabular}
\end{ruledtabular}
\end{minipage}
\end{table}

\section{Case Studies}\label{app:case-study}

We now apply these formulas to three search tasks
of increasing primitive complexity.  We reduce each task from a search
target to the inputs of \eqref{eq:twopass}: $M$, $N$, $N_e$; the numbers of spectral, static, and dynamical primitives
$N_{P,\mathrm{spec}}$, $N_{P,\mathrm{stat}}$, $N_{P,\mathrm{dyn}}$;
$\rho_{\mathrm{coarse}}$ via \eqref{eq:rho-decomp}; the spectral
gap~$\Delta$, instantiating the uniform gap assumption at the target's
physical isolation scale; and, where
dynamical primitives are active, the dynamical broadening~$\eta$ and
resolvent order~$r$.  All three cases share
$\epsilon_{\mathrm{coarse}} = 10^{-1}$~Ha,
$p_{\mathrm{coarse}} = p_{\mathrm{fine}} = 0.95$,
$\rho_{\mathrm{fine}} = 1$, and
the grid tolerance $\epsilon_{\mathrm{grid}} = \sqrt{\epsilon}$ on each
pass, from \eqref{eq:grid-rule}.  All tolerances are absolute total-energy values in
Hartree.  We quote the corresponding tolerances per atom for each case.  We take all
three as zero-temperature ground-state searches
in which every constraint threshold is a fixed external constant
(formation and stability energies referenced to elemental or
competing-phase chemical potentials).  We take each as an illustration
of the formulas.  We assume that the
representative cell ranges meet the fixed kinetic-energy ordering
condition, so the reported costs apply to such restricted
cell domains.

\begin{figure*}[t]
\centering
\includegraphics[width=\textwidth]{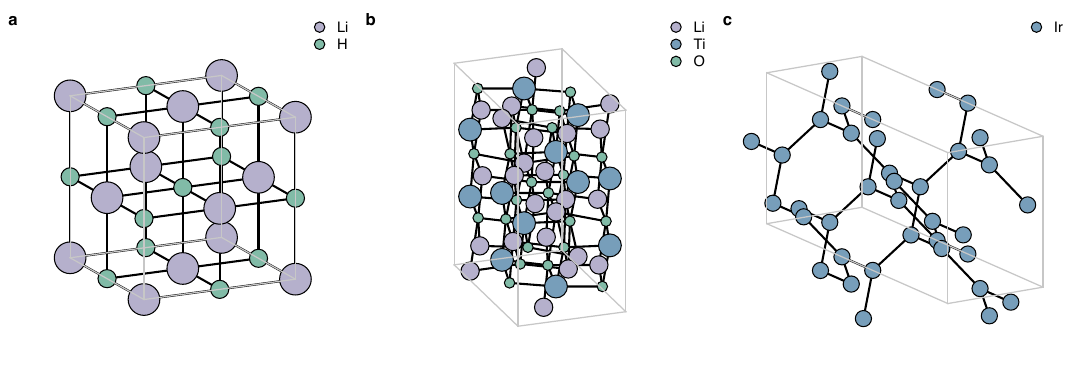}
\caption{Case-study materials.  Representative structures of the three
  targets, from experimental or DFT-relaxed data.  \textbf{a}, Stoichiometric lithium
  hydride: the rocksalt phase ($Fm\bar3m$, $a = 4.09$~\AA) recovered in the lithium-hydride case
  \cite{zurekLittleBitLithium2009}.  \textbf{b}, Solid-state tritium breeder: the
  $\beta$-Li$_2$TiO$_3$ host ($C2/c$, $a,b,c = 5.04, 8.81, 9.73$~\AA,
  $\beta = 100^\circ$) \cite{dorrianRefinementStructureLi2TiO31969}.
  \textbf{c}, Three-dimensional quantum spin liquid: a class represented
  here by the Ir hyperhoneycomb sublattice of $\beta$-Li$_2$IrO$_3$ ($Fddd$)
  \cite{MaterialsDataLi2IrO32020}.  Grey lines outline the unit cell.  In \textbf{c}, neighbouring Ir
  sites beyond the cell are included to complete the connected net.
  Ball radii are display values, bonds are distance-threshold lines,
  and each panel is scaled independently.}%
\label{fig:cases}
\end{figure*}

\emph{Stoichiometric lithium hydride.}
From the 1:1 Li:H stoichiometry alone the search ranges over bounded
two-atom primitive-cell geometries for the most thermodynamically stable
solid.  A configuration is a solution when its ground-state energy falls below
the constraint threshold.  The decoded solution configuration is the known
rocksalt lithium-hydride phase, with conventional cubic lattice constant $\approx4.09$~\AA\
\cite{zurekLittleBitLithium2009} (\cref{fig:cases}a).  Fixed composition gives $M=2$, $N=2$.
The number of valence electrons is $N_e=2$.
The single target is one zeroth-order spectral primitive:
$N_{P,\mathrm{spec}}=1$,
$N_{P,\mathrm{stat}}=N_{P,\mathrm{dyn}}=0$,
$N_P=1$, $N_{\mathrm{prop}}=1$, no $\eta$ or~$r$.
The gap is $\Delta = 0.1834$~Ha, the $4.99$~eV ambient-pressure band
gap~\cite{zurekLittleBitLithium2009}.
The verification tolerance $\epsilon_E = 10^{-2}$~Ha is an absolute
total-energy tolerance (${\approx}136$~meV per atom for this two-atom
cell), the coarsest tolerance of the three cases, comparable to the
${\sim}100$~meV/atom metastability windows used in convex-hull
screening~\cite{tolborgFreeEnergyPredictions2022}.
We take the coarse solution density
$\rho_{\mathrm{coarse}} = 10^{-2}$ as a lower bound on the low-energy region
fraction in the bounded two-atom ensemble.  With stoichiometry fixed, the
site-by-site number of species does not apply.
We obtain $Q_{\mathrm{case}}^{(2)}\sim 10^{2}$ logical qubits and
$D_{\mathrm{case}}^{(2)}\sim 10^{14}$ Toffoli depth.  The register is small
(two nuclei, two electrons, one spectral primitive), with depth set by
state isolation at the band gap in the fine pass and with no broadening or
resolvent factor.

\emph{Solid-state tritium breeder.}
The search ranges over bounded small-cell Li--Ti--O hosts carrying one
interstitial tritium probe.
We screen by combining (i) a formation-energy threshold,
(ii) a tritium diffusion-barrier limit along a hop path, and
(iii) thermal stability (no imaginary phonon modes, plus a
thermal-conductivity proxy).
Li$_2$TiO$_3$-type ceramics (\cref{fig:cases}b) are leading solid tritium breeders in which
tritium transport is governed by bulk and grain-boundary diffusion and
point-defect traps~\cite{xuRecentProgressResearch2022}, while
localized, correlated electronic states of oxide defects are, in general,
a known failure mode of semilocal DFT, which delocalizes them without a
Hubbard-$U$ or hybrid
correction~\cite{fallettaHubbardPolaronicDefect2022}.
The bounded class comprises the Li$_2$TiO$_3$-ratio-biased Li--Ti--O/T
small-cell compositions: $M=4$; one Li$_2$TiO$_3$ formula unit plus the
interstitial probe, $N=7$.  The number of HGH valence electrons is $N_e=25$
(Li~1, Ti~4, O~6, T/H~1).
Formation energy is one spectral primitive.  The diffusion barrier gives
one projected-displacement-force primitive (first-order static,
Hellmann--Feynman) at each point of a 5-point path
($N_{P,\mathrm{stat}}=5$).  We evaluate phonon stability and a harmonic
conductivity proxy on a retained grid of 24 phonon-response samples, one
dynamical finite-resolvent response each ($N_{P,\mathrm{dyn}}=24$).
Hence $N_{P,\mathrm{spec}}=1$, $N_{P,\mathrm{stat}}=5$,
$N_{P,\mathrm{dyn}}=24$, $N_P=30$, $N_{\mathrm{prop}}=3$.
The gap is $\Delta = 2\times10^{-2}$~Ha, a rounded $0.5$~eV polaronic
isolation scale motivated by the reported polaron formation and hopping
energetics~\cite{fallettaHubbardPolaronicDefect2022}.  With dynamical
primitives active, $r=1$, $\eta_{\mathrm{coarse}} = 4\times10^{-4}$~Ha, and
$\eta_{\mathrm{fine}} = 4\times 10^{-5}$~Ha (${\approx}1.1$~meV), below
typical phonon linewidths.
We set $\rho_{\mathrm{coarse}}=3.16\times10^{-4}$
with $M_{\mathrm{eff}}\approx 3.16$ over $N=7$ and one region-weighted
target.  We restrict the allowed species by the ratio bias to approximately three of the four at each site, with uniform preparation over the allowed species.  We use
$\epsilon_E = 2\times 10^{-2}$~Ha (${\approx}78$~meV per atom) in the fine pass, sufficient
to resolve the formation-energy and barrier pass/fail margins.
We obtain $Q_{\mathrm{case}}^{(2)}\sim 10^{5}$ logical qubits and
$D_{\mathrm{case}}^{(2)}\sim 10^{19}$ Toffoli depth.
Qubits are set by the pairwise distance and potential registers of the active primitives, with the total increased by
the 24 dynamical phonon primitives.  Depth is set by state isolation at the
polaronic gap.

\emph{Three-dimensional quantum spin liquid.}
The search ranges over heavy-element ternary/quaternary
configurations (spin-orbit-active $4d/5d$ or lanthanide magnetic cations
on frustrated networks) for a stable 3D topological quantum spin
liquid, a U(1) Coulomb quantum-spin-ice or a 3D Kitaev-like phase
(\cref{fig:cases}c).  We screen with
a finite, representative set of necessary signatures (a fixed
momentum--frequency grid for the dynamical response) that separate a
configuration from a trivial paramagnet or a conventionally ordered magnet,
short of confirming topological order:
(i) structural and thermodynamic stability (energy, forces, stress,
local curvature);
(ii) a robust insulating charge gap;
(iii) absence of local magnetic dipole order;
(iv) a fractionalized-excitation signature (broad continuum in the
dynamical spin response, not sharp magnon poles).
For 3D Kitaev spin liquids, quantum Monte Carlo's sign-free property is lost once non-Kitaev interactions or magnetic fields are present,
and purification-based thermal tensor networks converge slowly at the
low temperatures near the spin-liquid ground state owing to entanglement
growth~\cite{jahromiThermodynamicsThreedimensionalKitaev2021}.
We use $M=6$, $N=16$, a three-dimensional
frustrated-network cell.  The number of heavy-element HGH valence electrons
is $128$.  The electron-addition energy evaluated below raises the register to $N_e=129$.
Neutral and electron-addition/removal ground-state energies give
$N_{P,\mathrm{spec}}=3$.  Forces (48), stress (6), and local magnetic
moments (12, on four magnetic cation sites) are first-order static
primitives ($N_{P,\mathrm{stat}}=66$).  A spin-structure-factor grid ($4$
momenta $\times\, 2$ frequency windows $\times\, 3$ diagonal spin components)
gives higher-order dynamical responses ($N_{P,\mathrm{dyn}}=24$).
Hence $N_P=93$, $N_{\mathrm{prop}}=6$.
The gap and fine linewidth are set at the magnetic exchange scale,
$\Delta = \eta_{\mathrm{fine}} = 5\times 10^{-4}$~Ha (${\approx}13.6$~meV),
the 10--15~meV Kitaev exchange scale reported for hyperhoneycomb
$\beta$-Li$_2$IrO$_3$~\cite{katukuriVicinityHyperhoneycombVLi2IrO32016},
with $r=1$ and $\eta_{\mathrm{coarse}} = 10^{-3}$~Ha.  We set
$\rho_{\mathrm{coarse}}=10^{-4}$
with $M_{\mathrm{eff}}\approx 1.78$ over $N=16$ and one region-weighted
target.  Here the solution-density decomposition is only partly enumerable.
Fractionalization is an emergent many-body property, not a site-by-site one.
Consequently, $M_{\mathrm{eff}}$ here combines an enumerable structural
restriction on the candidate configurations with a non-enumerable rate at which
the allowed structures realize a quantum spin liquid.  The reported
$\rho_{\mathrm{coarse}}$ reflects both.  The pure site-by-site number of species
$M_{\mathrm{eff}}=M$, which assumes no structural prior, sets the corresponding upper
bound on the search factor.
We take the verification tolerance as $\epsilon_E = 10^{-2}$~Ha, ${\approx}17$~meV
per atom over the 16-atom cell, below chemical accuracy
($1\;\mathrm{kcal/mol}\approx43$~meV per atom), as required to resolve the fractionalization signatures.
We obtain $Q_{\mathrm{case}}^{(2)}\sim 10^{7}$ logical qubits and
$D_{\mathrm{case}}^{(2)}\sim 10^{26}$ Toffoli depth, dominated by state
isolation at the millielectronvolt magnetic gap of the heavy-element cell
and the assumed low solution density.

\bibliography{ref}

\end{document}